\documentclass{emulateapj}

\usepackage{graphicx, amsmath, amssymb, amsthm, rotating}
\usepackage{natbib}

\def\nhires{47}
\def\nuves{13}

\def\etal{et al.}

\def\lya{Ly$\alpha$ }

\def\mdm{$M_{\rm DM}$}

\def\kms{km~s$^{-1}$ }

\def\dv{${\Delta v}_{90}$}

\def\nh{$N_{\rm H I}$}

\def\w1526{$W_{1526}$}

\def\mcrit{$M_{\rm shock}$}
\def\mmed{$M_{\rm med}$}

\def\micron{$\mu$m}

\def\cm2{cm$^{-2}$}
\def\cm3{${\rm cm}^{-3}$}

\def\cmma{\;\;\; ,}

\def\Mdot{$\dot{M_{*}}$}

\def\msolar{M$_{\odot}$}

\def\ps{$\dot{\psi_{*}}$}

\def\ciis{C II$^{*}$}
\def\lc{${\ell}_{c}$}
\def\lcmax{${\ell}^{\rm max}_{c}$}

\def\lccrit{${\ell}^{\rm crit}_{c}$}
\def\delc{${\Delta {\rm log_{10}}{{\ell}^{\rm crit}_{c}}}$}
\def\lctrue{${\ell}^{\rm true}_{c}$}

\def\jnu{$J_{\nu}$}
\def\jnulocal{$J_{\nu}^{\rm local}$}

\def\jnubkd{$J_{\nu}^{\rm bkd}$}

\def\sigmasfr{${\Sigma_{\rm SFR}}$}

\def\lcunit{ergs s$^{-1}$ H$^{-1}$}
\def\sfrunit{M$_{\odot}$ yr$^{-1}$ kpc$^{-2}$}
\def\arcsec{\hbox{$^{\prime\prime}$}}

\begin{document}

\title{BIMODALITY IN  DAMPED {\lya} SYSTEMS}

\author{ Arthur M. Wolfe\altaffilmark{1,2}, 
Jason X. Prochaska\altaffilmark{1,3},
Regina A. Jorgenson\altaffilmark{1,2} 
\& Marc Rafelski\altaffilmark{1,2}}
\altaffiltext{1}{Visiting Astronomer, W.M. Keck Telescope.
The Keck Observatory is a joint facility of the University
of California and the California Institute of Technology.}
\altaffiltext{2}{
Department of Physics, and Center for Astrophysics and Space Sciences; 
University of California, San Diego; 
9500 Gilman Dr.; La Jolla; CA 92093-0424; {\tt awolfe@ucsd.edu,
regina@physics.ucsd.edu,marcar@ucla.edu}}
\altaffiltext{3}{
Department of Astronomy \& Astrophysics,  UCO/Lick Observatory;
1156 High Street;
University of California; Santa Cruz, CA 95064;
{\tt xavier@ucolick.org}}

\begin{abstract} 
We report evidence for a bimodality in 
 damped {\lya} systems (DLAs). Using [C II] 158
{\micron} cooling rates, {\lc}, 
we find a distribution with
peaks at {\lc}=10$^{-27.4}$ and 10$^{-26.6}$ {\lcunit} separated
by a trough at {\lccrit} $\approx$ 10$^{-27.0}$ {\lcunit}.
We divide the sample into `low cool' DLAs with
{\lc} $\le$ {\lccrit} and `high cool' DLAs with
{\lc} $>$ {\lccrit} and find the Kolmogorv-Smirnov probabilities
that velocity width, metallicity,
dust-to-gas ratio, and Si II equivalent width in the two
subsamples are drawn from the same parent population are small.
All these quantities are significantly larger in the `high cool'
population,
while the H I column densities are indistinguishable in the
two populations. 
We find that heating by X-ray and FUV background radiation is
insufficient to balance the cooling rates of either population.
Rather, the DLA gas is heated by local radiation fields.
The rare appearance
of faint, extended objects
in the Hubble Ultra Deep Field rules out {\it in situ}
star formation as the dominant star-formation mode
for the `high cool' population, but is 
compatible with {\it in situ} star formation as the
dominant mode for the `low cool'
population. Star formation in the `high cool' DLAs likely
arises in Lyman Break galaxies. 
We investigate
whether these properties of DLAs are analogous to the
bimodal properties of nearby galaxies. 
Using Si II equivalent width as a mass indicator,
we construct bivariate distributions of metallicity,
{\lc}, and areal SFR versus the mass indicators.
Tentative evidence is found for correlations and 
parallel sequences, which suggest similarities between
DLAs and nearby galaxies.
We suggest that the transition-mass model
provides
a plausible scenario for the bimodality we have found.
As a result, the bimodality in current
galaxies may have originated in DLAs.

\end{abstract}


\section{INTRODUCTION}

Recent studies of $\sim$ 10$^{5}$ low-redshift galaxies
reveal a striking bimodality in their properties. The
galaxy population is divided into two mass sequences; 
a `blue' sequence with stellar masses $M_{*}$ $<$
10$^{10.5}$ {\msolar} and a `red' sequence
in which  
$M_{*}$ $>$
10$^{10.5}$ {\msolar} (Kauffmann {\etal} 2003;
Baldry {\etal} 2004). The
`blue' sequence is comprised of late-type galaxies undergoing
active star formation, while the `red' sequence consists
of early-type galaxies with little, if any, star formation.
Dekel \& Birnboim (2006) suggest that the mass sequences
reflect the different paths for the build up of stellar populations
in galaxies. The critical parameter in their scenario is the
dark-matter halo `shock mass' , $M_{\rm shock}$=10$^{11.5}$ {\msolar},
which corresponds to
$M_{*}$ = 
10$^{10.5}$ {\msolar}.
At high redshifts, gas accreted onto
halos
with masses {\mdm} $<$ $M_{\rm shock}$
is not heated to the virial temperature of the halo, but rather
produces 
star-forming disks through the infall of cold streams. The continuation
of this process to the present
results in the formation of the `blue' sequence. 
By contrast
gas accreted onto
halos with masses {\mdm} $>$ $M_{\rm shock}$
is shock
heated to the virial temperature, and accretes onto
the halos in a hot  cooling flow. Cold
filaments penetrate the hot gas
and their inflow
results in star-forming bulges in massive galaxies
(Dekel \& Birnboim 2006; see also Bell {\etal} 2004; Faber
{\etal} 2007). 
In this case the low density of the hot gas
and feedback processes
suppress star formation  at $z$ $\le$ 2. As a result  these halos evolve
onto the `red' sequence. 

Our purpose here  is to present evidence 
for an analogous bimodality in the properties of damped {\lya} 
systems (DLAs). 
Because these neutral-gas layers
are the likely progenitors of modern galaxies
(see Wolfe, Gawiser, \& Prochaska 2005 [hereafter referred
to as WGP05] for a review),
the  DLA bimodality may be related to the galaxy
phenomenon discussed above. In fact we shall argue that the
bimodality in modern galaxies originates in DLAs. 
We use a technique which, for the first time, measures 
bimodality in absorption-line gas \footnote {Prochaska {\etal} (2002)
report evidence for bimodality in the ratio
of the nitrogen to alpha element abundance, but its statistical
significance is tentative owing to the small size of the sample.}.  
Rather than obtain signatures based
on starlight emitted by galaxies (e.g. Kauffmann {\etal} 2003), we rely
on an absorption-line diagnostic, which we argue is
a signature of star formation.
Specifically, we measure the
{\ciis} $\lambda $1335.7 absorption line, which,
if thermal balance is assumed,
is an indicator of the rate at which neutral gas is heated.
The crucial parameter is
the
[C II] 158 {\micron} cooling rate per H atom, {\lc},
which divides the DLA sample in such a way that
objects  with {\lc} $>$ {\lccrit} differ
fundamentally from those with {\lc} $\le$ 
{\lccrit}, where {\lccrit}
$\sim$10$^{-27}$ {\lcunit}. We shall argue that
{\lc} is a tracer of star formation rates (see
Wolfe, Prochaska, \& Gawiser 2003 [hereafter referred 
to as WPG03]) and that the
bimodality in {\lc} is physically related
to a transition in star-formation modes, from
{\it in situ} star formation in DLAs with 
{\lc} $\le$ {\lccrit} to star formation in
compact `bulge' regions sequestered away from
the DLA gas in objects in which {\lc} $>$ {\lccrit}:
the higher star formation rates (SFRs) 
predicted for the `bulge' models
account for the higher heating rates of the surrounding
DLA gas
(see $\S$ 7).
We shall argue that the bimodality in 
star formation modes is caused by a
transition in galaxy-formation modes, which ultimately is due
to  a transition in mass 
(Dekel \& Birnboim (2006).

The paper is organized as follows. 
In $\S$ 2 we examine the {\lc}
distribution and discuss the results of tests to assess
whether the distribution is bimodal. In $\S$ 3
we divide our DLA sample into two sub-samples: `low cool' DLAs 
with {\lc} $\le$ {\lccrit} and `high cool' DLAs  with {\lc} $>$
{\lccrit}. For each sub-sample we compile distributions
of physical parameters such as absorption-line velocity width,
metallicity, dust-to-gas ratio, Si II $\lambda$ 1526
equivalent width, and H I column density {\nh}. We then 
determine the probability that the pair of distributions
corresponding
to a given parameter is
drawn
from the same parent population, and describe the results
of these tests. In $\S$ 4 we discuss
the physical significance
of {\lccrit} and conclude that it signifies a dividing line
between two modes of star formation. In $\S$ 5 we give a brief summary
of the results and conclude that the crucial parameter distinguishing
the two populations is dark-matter mass. In $\S$ 6 we draw analogies
between bimodality in DLAs and in modern galaxies and show how the
two are related. We also place these results in the context
of modern theories of galaxy formation. Conclusions
are given in $\S$ 7.

Throughout this paper we adopt a  cosmology with
(${\Omega_{\rm M}},{\Omega_{\Lambda}},h$)=(0.3,0.7.0.7).

\begin{sidewaystable*}\footnotesize
\begin{center}
\caption{\sc {DLA Sample\label{tab:stat}}}
\begin{tabular}{lcccccccccccc}
\tableline
\tableline
Quasar &
RA (2000) & 
DEC (2000) &
$z_{abs}^a$ &
log$_{10} N_{\rm HI}^b$ & 
$\log_{10} N({\rm CII}^*)^c$ & 
$\log_{10} \ell_c^{d}$ &
$\Delta v_{90}^{e}$ &
[M/H]$^{f}$ & 
[Fe/H]$^{g}$ &
$W_{\rm 1526}^{h}$ & 
Ref \\
\tableline
Q0405-443&00:00:00.00&00:00:00.00&2.5950&$20.90^{+0.10}_{-0.10}$&$ 13.66\pm0.21$&$-26.76\pm0.23$&  79&$-0.96\pm0.10$&$-1.33\pm0.10$&---&24\\
Q2359-02&00:01:50.00&-01:59:40.34&2.1539&$20.30^{+0.10}_{-0.10}$&$< 14.48$&$<-25.34$&  78&$-1.58\pm0.01$&$-1.88\pm0.03$&0.28&6,13\\
Q2359-02&00:01:50.00&-01:59:40.34&2.0951&$20.70^{+0.10}_{-0.10}$&$ 13.70\pm0.06$&$-26.51\pm0.12$& 142&$-0.77\pm0.02$&$-1.65\pm0.03$&0.83&6,13\\
J001328.21+135827&00:13:28.21&+13:58:27.9&3.2811&$21.55^{+0.15}_{-0.15}$&$ 13.67\pm0.04$&$-27.40\pm0.16$&  10&$-2.10\pm0.16$&$-2.72\pm0.02$&0.28&32\\
BR0019-15&00:22:08.01&-15:05:38.78&3.4389&$20.92^{+0.10}_{-0.10}$&$ 13.84\pm0.02$&$-26.60\pm0.10$& 118&$-1.06\pm0.05$&$-1.58\pm0.04$&0.76&6,13\\
PH957&01:03:11.38&+13:16:16.7&2.3090&$21.37^{+0.08}_{-0.08}$&$ 13.59\pm0.05$&$-27.30\pm0.09$&  56&$-1.46\pm0.01$&$-1.90\pm0.04$&0.38&1,6,13\\
SDSS0127-00&01:27:00.69&-00:45:59&3.7274&$21.15^{+0.10}_{-0.10}$&$ 13.20\pm0.06$&$-27.46\pm0.12$&  40&$-2.40\pm0.10$&$-2.90\pm0.02$&0.28&22\\
PSS0133+0400&01:33:40.4&+04:00:59&3.7736&$20.55^{+0.10}_{-0.15}$&$ 14.02\pm0.01$&$-26.05\pm0.10$& 123&$-0.75\pm0.10$&$-1.07\pm0.08$&0.92&22\\
PSS0133+0400&01:33:40.4&+04:00:59&3.6919&$20.70^{+0.10}_{-0.15}$&$ 12.95\pm0.03$&$-27.27\pm0.11$&  39&$-2.34\pm0.15$&$-2.74\pm0.05$&0.31&22\\
J013901.40-082443&01:39:01.40&-08:24:43.9&2.6773&$20.70^{+0.15}_{-0.15}$&$ 13.81\pm0.03$&$-26.40\pm0.15$& 110&$-1.27\pm0.19$&$-1.62\pm0.02$&0.67&32\\
Q0149+33&01:52:34.472&+33:50:33.23&2.1408&$20.50^{+0.10}_{-0.10}$&$< 12.78$&$<-27.24$&  40&$-1.49\pm0.05$&$-1.77\pm0.02$&0.22&6,13\\
Q0201+11&02:03:46.53&11:34:40.4&3.3869&$21.26^{+0.10}_{-0.10}$&$ 14.12\pm0.10$&$-26.66\pm0.14$&  67&$-1.25\pm0.11$&$-1.41\pm0.05$&0.64&15\\
PSS0209+0517&02:09:44.52&+05:17:17.3&3.8636&$20.55^{+0.10}_{-0.10}$&$< 12.51$&$<-27.55$&  47&$-2.60\pm0.10$&$-2.96\pm0.09$&0.06&22\\
SDSS0225+0054&02:25:54.85&+00:54:51&2.7137&$21.00^{+0.15}_{-0.15}$&$> 13.37$&$>-27.15$&  60&$-0.91\pm0.14$&$-1.31\pm0.04$&1.08&29\\
J023408.97-075107&02:34:08.97&-07:51:07.6&2.3180&$20.95^{+0.15}_{-0.15}$&$< 13.41$&$<-27.05$&  10&$-2.74\pm0.14$&$-3.14\pm0.04$&0.09&32\\
J0255+00&02:55:18.62&+00:48:47.94&3.9146&$21.30^{+0.05}_{-0.05}$&$ 13.44\pm0.04$&$-27.37\pm0.06$&  38&$-1.78\pm0.01$&$-2.05\pm0.09$&---&13\\
J0307-4945&03:07:22.85&-49:45:47.6&4.4679&$20.67^{+0.09}_{-0.09}$&$< 13.59$&$<-26.60$& 192&$-1.55\pm0.08$&$-1.96\pm0.21$&---&11\\
Q0336-01&03:39:00.99&-01:33:18.07&3.0621&$21.20^{+0.10}_{-0.10}$&$ 14.01\pm0.01$&$-26.71\pm0.10$& 102&$-1.54\pm0.01$&$-1.81\pm0.02$&---&13\\
Q0347-38&03:49:43.54&-38:10:04.91&3.0247&$20.63^{+0.00}_{-0.00}$&$ 13.47\pm0.03$&$-26.68\pm0.03$&  84&$-1.16\pm0.03$&$-1.62\pm0.01$&0.45&6,13,20\\
Q0458-02&05:01:12.81&-01:59:14.25&2.0396&$21.65^{+0.09}_{-0.09}$&$> 14.80$&$>-26.37$&  82&$-1.19\pm0.02$&$-1.76\pm0.05$&0.67&6,13\\
HS0741+4741&07:45:21.75&+47:34:35.56&3.0174&$20.48^{+0.10}_{-0.10}$&$< 12.55$&$<-27.44$&  42&$-1.68\pm0.00$&$-1.93\pm0.00$&0.21&13\\
FJ0747+2739&07:47:11.19&+27:39:03.6&3.9000&$20.50^{+0.10}_{-0.10}$&$ 13.35\pm0.07$&$-26.66\pm0.12$& 150&$-2.01\pm0.01$&$-2.45\pm0.03$&0.22&22\\
FJ0812+32&08:12:40.8&+32:08:08&2.6263&$21.35^{+0.10}_{-0.10}$&$ 14.30\pm0.01$&$-26.56\pm0.10$&  70&$-0.93\pm0.05$&$-1.76\pm0.01$&0.60&22,31,\\
J081435.18+502946&08:14:35.18&+50:29:46.5&3.7082&$21.35^{+0.15}_{-0.15}$&$< 13.18$&$<-27.69$&  30&$-3.00\pm0.15$&$-2.91\pm0.03$&0.25&32\\
J082619.70+314848&08:26:19.70&+31:48:48.0&2.9122&$20.30^{+0.15}_{-0.15}$&$< 12.60$&$<-27.21$&  35&$-1.88\pm0.15$&$-2.01\pm0.01$&---&32\\
Q0836+11&08:39:33.015&+11:12:03.82&2.4653&$20.58^{+0.10}_{-0.10}$&$< 13.12$&$<-26.97$&  88&$-1.15\pm0.05$&$-1.40\pm0.01$&0.59&13\\
J0929+2825&09:29:14.49&+28:25:29.1&3.2627&$21.10^{+0.00}_{-0.00}$&$ 13.15\pm0.02$&$-27.47\pm0.02$&  43&$-1.62\pm0.01$&$-1.78\pm0.01$&---&32\\
BR0951-04&09:53:55.69&-05:04:18.5&4.2029&$20.40^{+0.10}_{-0.10}$&$ 13.37\pm0.08$&$-26.54\pm0.13$&  36&$-2.62\pm0.03$&$<-2.57$&0.04&6,13\\
BRI0952-01&09:55:00.10&-01:30:06.94&4.0244&$20.55^{+0.10}_{-0.10}$&$ 13.55\pm0.02$&$-26.52\pm0.10$& 302&$-1.46\pm0.18$&$-1.86\pm0.08$&---&8,13\\
PC0953+47&09:56:25.2&+47:34:44&4.2442&$20.90^{+0.15}_{-0.15}$&$ 13.60\pm0.10$&$-26.82\pm0.18$&  70&$-2.19\pm0.03$&$-2.52\pm0.08$&0.24&22\\
J1014+4300&10:14:47.18&+43:00:30.1&2.9588&$20.50^{+0.00}_{-0.00}$&$ 12.76\pm0.04$&$-27.25\pm0.04$& 108&$-0.71\pm0.14$&$-1.11\pm0.04$&---&32\\
Q1021+30&10:21:56.84&+30:01:31.3&2.9489&$20.70^{+0.10}_{-0.10}$&$< 12.91$&$<-27.31$& 100&$-1.94\pm0.02$&$-2.16\pm0.01$&0.25&13,22\\
J103514.22+544040&10:35:14.22&+54:40:40.1&2.6840&$20.50^{+0.20}_{-0.20}$&$> 13.59$&$>-26.42$& 160&$-0.62\pm0.21$&$-0.45\pm0.12$&1.43&32\\
Q1036-230&10:39:09.4&-23:13:26&2.7775&$21.00^{+0.10}_{-0.10}$&$ 12.97\pm0.05$&$-27.55\pm0.11$&  80&$-1.41\pm0.10$&$-1.82\pm0.10$&---&13\\
Q1104-18&11:06:32.96&-18:21:09.82&1.6614&$20.80^{+0.10}_{-0.10}$&$ 13.44\pm0.05$&$-26.88\pm0.11$&  56&$-1.04\pm0.01$&$-1.48\pm0.02$&---&7\\
BRI1108-07&11:11:13.64&-08:04:02.47&3.6076&$20.50^{+0.10}_{-0.10}$&$< 12.34$&$<-27.67$&  32&$-1.80\pm0.00$&$-2.12\pm0.01$&0.19&8,13\\
J113130.41+604420&11:31:30.41&+60:44:20.7&2.8760&$20.50^{+0.15}_{-0.15}$&$< 12.66$&$<-27.36$&  53&$-2.13\pm0.15$&$-2.31\pm0.03$&---&32\\
HS1132+2243&11:35:08.03&+22:27:06.8&2.7835&$21.00^{+0.07}_{-0.07}$&$< 12.69$&$<-27.82$&  56&$-1.99\pm0.07$&$-2.22\pm0.02$&0.27&22\\
J115538.60+053050&11:55:38.60&+05:30:50.6&3.3268&$21.05^{+0.10}_{-0.10}$&$ 13.73\pm0.03$&$-26.84\pm0.10$& 120&$-0.81\pm0.10$&$-1.44\pm0.05$&1.21&32\\
Q1157+014&11:59:44.81&+01:12:07.1&1.9440&$21.80^{+0.10}_{-0.10}$&$> 14.80$&$>-26.51$&  84&$-1.36\pm0.06$&$-1.81\pm0.04$&0.73&9\\
BR1202-07&12:05:23.63&-07:42:29.91&4.3829&$20.60^{+0.14}_{-0.14}$&$< 13.06$&$<-27.06$& 170&$-1.81\pm0.02$&$-2.19\pm0.12$&---&2\\
J120802.65+630328&12:08:02.65&+63:03:28.7&2.4439&$20.70^{+0.15}_{-0.15}$&$ 13.55\pm0.03$&$-26.67\pm0.15$&  60&$-2.32\pm0.15$&$-2.55\pm0.01$&0.18&32\\
Q1215+33&12:17:32.54&+33:05:38.39&1.9991&$20.95^{+0.07}_{-0.07}$&$< 13.17$&$<-27.29$&  42&$-1.48\pm0.03$&$-1.70\pm0.05$&0.35&6,13\\
Q1223+17&12:26:07.22&+17:36:48.98&2.4661&$21.50^{+0.10}_{-0.10}$&$< 14.01$&$<-27.01$&  94&$-1.59\pm0.01$&$-1.84\pm0.02$&0.62&8,13\\
Q1232+08&12:34:37.55&+07:58:40.5&2.3371&$20.90^{+0.10}_{-0.10}$&$ 14.00\pm0.10$&$-26.42\pm0.14$&  85&$-1.28\pm0.09$&$-1.72\pm0.09$&---&9\\
J1240+1455&12:40:20.91&+14:55:35.6&3.1078&$21.30^{+0.00}_{-0.00}$&$> 14.34$&$>-26.47$& 335&$-0.85\pm0.03$&$-2.11\pm0.02$&---&32\\
J1240+1455&12:40:20.91&+14:55:35.6&3.0241&$20.45^{+0.00}_{-0.00}$&$< 13.32$&$<-26.65$& 134&$-0.74\pm0.07$&$-0.81\pm0.12$&---&32\\
Q1331+17&13:33:35.78&+16:49:04.03&1.7764&$21.14^{+0.08}_{-0.08}$&$< 13.54$&$<-27.12$&  72&$-1.42\pm0.00$&$-2.02\pm0.00$&0.50&6,13\\
\tableline
\end{tabular}
\end{center}
\end{sidewaystable*}

\begin{sidewaystable*}\footnotesize
\begin{center}
\begin{tabular}{lcccccccccccc}
\tableline
\tableline
Quasar &
RA (2000) & 
DEC (2000) &
$z_{abs}^a$ &
log$_{10} N_{\rm HI}^b$ & 
$\log_{10} N({\rm CII}^*)^c$ & 
$\log_{10} \ell_c^{d}$ &
$\Delta v_{90}^{e}$ &
[M/H]$^{f}$ & 
[Fe/H]$^{g}$ &
$W_{\rm 1526}^{h}$ & 
Ref \\
\tableline
Q1337+11&13:40:02.44&+11:06:29.6&2.7959&$20.95^{+0.10}_{-0.10}$&$ 13.11\pm0.10$&$-27.36\pm0.14$&  60&$-1.72\pm0.12$&$-2.03\pm0.08$&0.25&22,31\\
BRI1346-03&13:49:16.82&-03:37:15.06&3.7358&$20.72^{+0.10}_{-0.10}$&$ 12.55\pm0.11$&$-27.69\pm0.15$&  38&$-2.33\pm0.01$&$-2.63\pm0.02$&0.12&6,13\\
PKS1354-17&13:57:06.07&-17:44:01.9&2.7800&$20.30^{+0.15}_{-0.15}$&$ 12.76\pm0.06$&$-27.06\pm0.16$&  30&$-1.37\pm0.19$&$-1.79\pm0.05$&0.21&22\\
J141030+511113&14:10:30.60&+51:11:13.5&2.9642&$20.85^{+0.20}_{-0.20}$&$< 13.01$&$<-27.35$&  46&$-1.96\pm0.15$&$-2.27\pm0.02$&---&\\
J141030+511113&14:10:30.60&+51:11:13.5&2.9344&$20.80^{+0.15}_{-0.15}$&$> 13.39$&$>-26.92$& 247&$-0.95\pm0.15$&$-1.16\pm0.10$&---&32\\
J141906.32+592312&14:19:06.32&+59:23:12.3&2.2476&$20.95^{+0.20}_{-0.20}$&$< 13.08$&$<-27.39$&  20&$-2.85\pm0.20$&$-2.76\pm0.04$&0.09&32\\
Q1425+6039&14:26:56.44&60:25:42.74&2.8268&$20.30^{+0.04}_{-0.04}$&$< 13.33$&$<-26.49$& 136&$-0.79\pm0.04$&$-1.32\pm0.00$&0.70&2,13,31\\
PSS1443+27&14:43:31.22&+27:24:37.23&4.2241&$20.80^{+0.10}_{-0.10}$&$> 14.71$&$>-25.61$&  90&$-0.70\pm0.16$&$-1.10\pm0.06$&---&8,13\\
PSS1506+5220&15:06:54.6&+52:20:05&3.2244&$20.67^{+0.07}_{-0.07}$&$< 12.91$&$<-27.28$&  44&$-2.98\pm0.08$&$-2.60\pm0.04$&0.16&22\\
Q1759+75&17:57:46.39&+75:39:16.01&2.6253&$20.76^{+0.01}_{-0.01}$&$ 12.80\pm0.05$&$-27.48\pm0.05$&  74&$-0.79\pm0.01$&$-1.18\pm0.00$&0.63&6,13\\
J203642.29-055300&20:36:42.29&-05:53:00.2&2.2805&$21.20^{+0.15}_{-0.15}$&$ 13.36\pm0.08$&$-27.36\pm0.17$&  71&$-1.71\pm0.17$&$-2.24\pm0.02$&0.31&32\\
SDSS2100-0641&21:00:25.03&-06:41:46&3.0924&$21.05^{+0.15}_{-0.15}$&$ 14.06\pm0.01$&$-26.51\pm0.15$& 187&$-0.73\pm0.15$&$-1.20\pm0.02$&---&29\\
J214129.38+111958&21:41:29.38&+11:19:58.3&2.4264&$20.30^{+0.20}_{-0.20}$&$< 13.28$&$<-26.54$&  30&$-1.97\pm0.20$&$-2.00\pm0.03$&0.15&32\\
J215117.00-070753&21:51:17.00&-07:07:53.3&2.3274&$20.45^{+0.15}_{-0.15}$&$< 13.09$&$<-26.87$&  20&$-1.65\pm0.15$&$-1.94\pm0.02$&0.31&32\\
Q2206-19&22:08:52.05&-19:43:57.61&1.9200&$20.65^{+0.07}_{-0.07}$&$ 13.97\pm0.25$&$-26.20\pm0.26$& 132&$-0.42\pm0.00$&$-0.86\pm0.02$&0.99&4,6,13\\
Q2206-19&22:08:52.05&-19:43:57.61&2.0762&$20.43^{+0.06}_{-0.06}$&$< 13.16$&$<-26.79$&  26&$-2.31\pm0.04$&$-2.61\pm0.02$&---&4,6,13\\
Q2231-002&22:34:08.80&+00:00:02.00&2.0661&$20.56^{+0.10}_{-0.10}$&$ 13.71\pm0.04$&$-26.37\pm0.11$& 122&$-0.88\pm0.02$&$-1.40\pm0.07$&0.79&2,6,13,25\\
J223438.52+005730&22:34:38.52&+00:57:30.0&2.8175&$20.80^{+0.20}_{-0.20}$&$< 13.50$&$<-26.81$&  80&$-0.99\pm0.21$&$-1.52\pm0.02$&0.66&32\\
J223843.56+001647&22:38:43.56&+00:16:47.9&3.3654&$20.40^{+0.15}_{-0.15}$&$< 13.12$&$<-26.80$&  30&$-2.34\pm0.15$&$-2.57\pm0.14$&0.09&32\\
BR2237-0607&22:39:53.39&-05:52:20.78&4.0803&$20.52^{+0.11}_{-0.11}$&$< 12.53$&$<-27.51$& 144&$-1.87\pm0.02$&$-2.14\pm0.12$&---&2\\
J231543.56+145606&23:15:43.56&+14:56:06.4&3.2729&$20.30^{+0.15}_{-0.15}$&$ 13.55\pm0.08$&$-26.27\pm0.17$& 110&$-1.78\pm0.15$&$-2.03\pm0.03$&0.37&32\\
FJ2334-09&23:34:46.44&-09:08:11.8&3.0569&$20.45^{+0.10}_{-0.10}$&$< 12.82$&$<-27.15$& 134&$-1.04\pm0.11$&$-1.49\pm0.01$&---&22\\
J2340-00&23:40:23.7&-00:53:27.0&2.0545&$20.35^{+0.15}_{-0.15}$&$ 13.84\pm0.04$&$-26.03\pm0.15$& 104&$-0.74\pm0.16$&$-0.92\pm0.03$&0.75&29,\\
J234352.62+141014&23:43:52.62&+14:10:14.6&2.6768&$20.50^{+0.15}_{-0.15}$&$< 12.96$&$<-27.06$&  35&$-1.50\pm0.28$&$-1.09\pm0.13$&---&32\\
Q2342+34&23:44:51.10&+34:33:46.8&2.9082&$21.10^{+0.10}_{-0.10}$&$ 13.70\pm0.06$&$-26.92\pm0.12$& 100&$-1.04\pm0.02$&$-1.58\pm0.06$&0.76&22,31\\
Q2343+125&23:46:28.22&+12:48:59.9&2.4313&$20.34^{+0.10}_{-0.10}$&$ 12.77\pm0.05$&$-27.09\pm0.11$& 290&$-0.54\pm0.01$&$-1.20\pm0.00$&---&5\\
Q2344+12&23:46:45.79&+12:45:29.98&2.5379&$20.36^{+0.10}_{-0.10}$&$< 12.95$&$<-26.93$&  66&$-1.74\pm0.01$&$-1.82\pm0.03$&---&2,13\\
Q2348-14&23:51:29.91&-14:27:47.55&2.2794&$20.56^{+0.08}_{-0.08}$&$< 13.21$&$<-26.87$&  30&$-1.92\pm0.02$&$-2.24\pm0.02$&0.19&6,13\\
\tableline
\end{tabular}
\end{center}
\tablenotetext{a}{ \ DLA redshift}
\tablenotetext{b}{ \ H I column density [cm$^{-2}$]}
\tablenotetext{c}{ \ {\ciis} column density [cm$^{-2}$]}
\tablenotetext{d}{ \ 158 {\micron} cooling rate per atom [{\lcunit}].}
\tablenotetext{e}{ \ Low-ion velocity width (km/s) as defined in \citep{pw97}.}
\tablenotetext{f}{ \ Logarithmic $\alpha$-metal abundance with respect to solar.}
\tablenotetext{g}{ \ Logarithmic Fe abundance with respect to solar.}
\tablenotetext{h}{ \ Rest frame Si II $\lambda$ 1526 equivalent width [{\AA}].}
\tablerefs{
1: \cite{wolfe94};
2: \cite{lu96};
3: \cite{pw96};
4: \cite{pw97b};
5: \cite{lu99};
6: \cite{pw99};
7: \cite{lopez99};
8: \cite{pw00};
9: \cite{petit00};
10: \cite{molaro00};
11: \cite{mirka01};
12: \cite{molaro01};
13: \cite{pro01};
14: \cite{pgw01};
15: \cite{eps+01};
16: \cite{pho+02};
17: \cite{lsp02};
18: \cite{lrd+02};
19: \cite{ldd+02};
20: \cite{lopez03};
21: \cite{songaila02};
22: \cite{p03_esi};
23: \cite{pro03b};
24: \cite{ledoux03};
25: \cite{mirka04};
26: \cite{ledoux06};
27: \cite{akerman05};
28: \cite{obp+06};
29: \cite{shf06};
30: \cite{dz06};
31: \cite{pwh+07};
32: This paper }
 
\end{sidewaystable*}

\begin{figure}[ht]
\figurenum{1}
\includegraphics[height=3.5in,angle=-90]{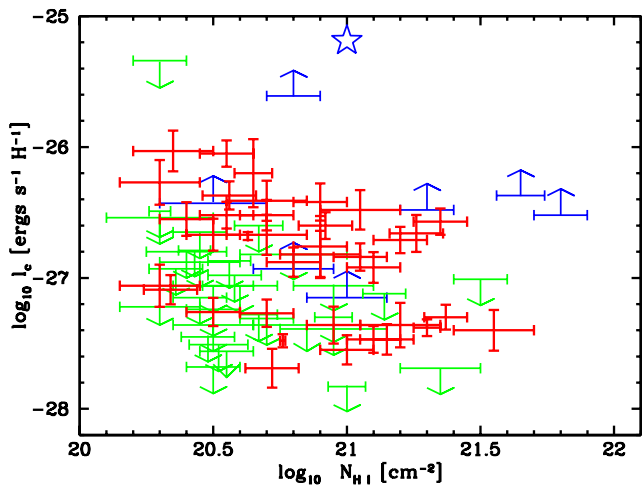}
\caption[]{{\lc} versus {\nh} for the 76 DLAs in our
sample. Red data points are positive detections, green
are upper limits, and blue are lower limits. 
The blue star is the average value for the  Galaxy disk.}
\label{fig:lcv_NHI}
\end{figure}

\section{BIMODALITY OF THE {\lc} DISTRIBUTION}

In this section we describe evidence for bimodality
in the distribution of 158 {\micron} cooling rates, {\lc}.
The values of {\lc} for the 76 DLAs in our sample are
listed in Table 1 (column 7).
As in
previous papers (e.g. WPG03)
we define {\lc} as follows:

\begin{equation}
{\ell}_{c}{\equiv} {{N({\rm CII^{*}})} \over {N_{\rm H I}}}A_{ul}h{\nu}_{ul}
\cmma
\end{equation}

\noindent where $N$({\ciis}) is the column density of the
excited $^{2}P_{3/2}$ state in the 2s$^{2}$2p term of C$^{+}$, $A_{ul}$
is the Einstein coefficient for spontaneous photon decay to
the ground $^{2}P_{1/2}$ state ($A_{ul}$=2.4{$\times$}10$^{-6}$ s$^{-1}$),
and 
$h{\nu_{ul}}$ is the
energy of this transition 
($h{\nu_{ul}}/k$=92 K).
All of the $N$({\ciis}) values in Table 1 (column 6) were deduced from
velocity profiles of the {\ciis} $\lambda$ 1335.7 transition 
arising from the $^{2}P_{3/2}$ state.  We
obtained \nhires\ profiles 
with the High Resolution Echelle Spectrometer (HIRES; Vogt {\etal} 1994) 
on the Keck  I 10 m telescope. The HIRES spectra were 
acquired using either a 0.8 {\arcsec} or 1.1 {\arcsec}
wide decker, resulting in velocity resolution
with FWHM = 6 and 8 {\kms} respectively.
Sixteen of the profiles were obtained with
the Echellette Spectrograph Imager
(ESI; Sheinis {\etal} 2002) on Keck II. The
ESI spectra were obtained with a 
0.5 {\arcsec} or 0.75 {\arcsec} slit (FWHM $\approx$ 
33 and 44 {\kms} respectively). 
The remaining \nuves\ profiles were obtained by other observers
with the UVES spectrograph on the VLT 8 m telescope.
The {\nh} values were
derived by fitting Voigt profiles to damped {\lya} lines detected
in the SDSS survey (Prochaska {\etal} 2005) and with
the Keck telescopes, the VLT, and several 4 m class
telescopes (Peroux {\etal} 2003; Storrie-Lombardi
\& Wolfe 2000; Wolfe {\etal} 1995).
The sample in Table 1 
comprises 32 upper limits (95 $\%$ confidence
level), 37 positive detections, and 7
lower limits on {\lc}. 

\begin{figure}
\figurenum{2}
\includegraphics[width=3.5in]{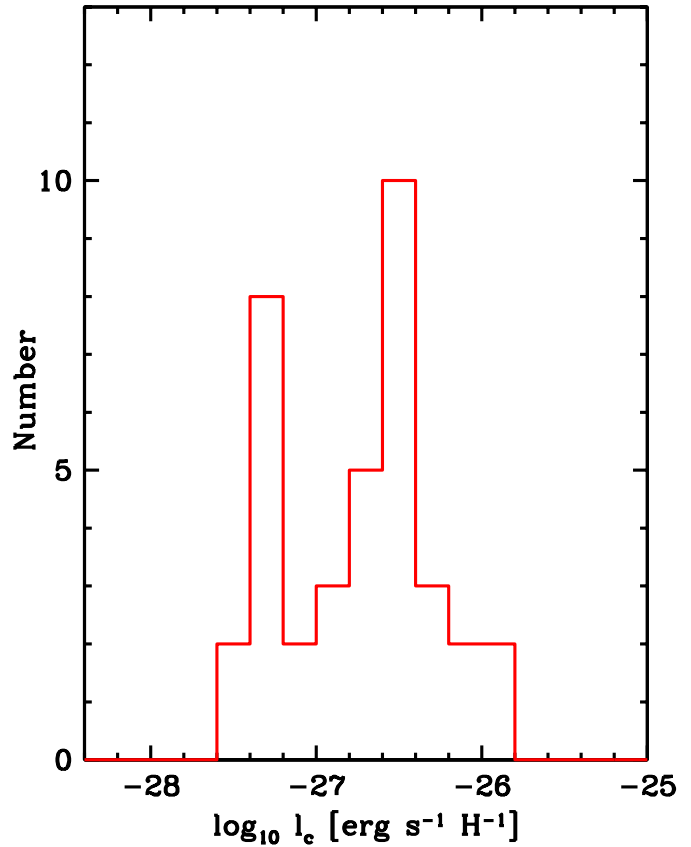}
\caption[]{Histogram depicting distribution
of the 37 positive detections of {\lc} reported in Table 1.}
\label{fig:lchist_pd}
\end{figure}

We first noticed evidence for bimodality in previous
plots of {\lc} versus
{\nh} (e.g.
WGP05). An updated version of these plots is shown in
Fig.~{\ref{fig:lcv_NHI}}. The red data points are positive
detections, green are upper limits, and blue are lower limits.
The positive detections are distributed into two distinct
regions centered near
{\lc}=10$^{-27.4}$ 
{\lcunit} and 
{\lc}=10$^{-26.6}$ {\lcunit}. A histogram of the positive detections
is plotted in  
Fig.~{\ref{fig:lchist_pd}}, which shows the two peaks 
separated by a trough at
{\lc} = {\lccrit}$\approx$10$^{-27.0}$
{\lcunit}. Note, the trough is not an artifact caused
by selection effects, since the signal-to-noise ratios
of the {\ciis} absorption profiles for the 5 DLAs with
{\lc}$\approx${\lccrit} are not exceptional for DLAs with
positive detections.
Standard statistical tests applied to these data 
support the bimodal hypothesis. 
We 
first used the KMM algorithm
(Ashman {\etal} 1994; McLachlan \& Basford 1987),
which fits Gaussians to the data,
computes the maximum likelihood estimates of their
means and variances, and evaluates the improvement of that fit over the null
hypothesis of a single Gaussian. We found the probability
of the null hypothesis to be $P_{\rm KMM}$({\lc})=0.016; i.e., a single
Gaussian fit can be excluded with 98.4 $\%$ confidence. This result
is robust, as the value of $P_{\rm KMM}$({\lc}) 
is insensitive to a wide range of
initial guesses for the means and variances of the input
Gaussians. We then used the Bayesian mixture algorithm
NMIX (Richardson \& Green 1997), which models the number of Gaussian
components and mixture component parameters jointly
and evaluates the statistical significance of these quantities
based on their posterior probabilities. J. Strader (priv. comm.; 2007) kindly
analyzed the data with NMIX and found that the
probability of the null hypothesis, $P_{\rm NMIX}$=0.08. 
The latter test
yielded peak locations of {\lc}=10$^{-27.34 {\pm}0.06}$ and
{\lc}=10$^{-26.58{\pm}0.06}$ {\lcunit} and values of 0.38 and 
0.62 for the sample fractions associated with the
respective peaks.

\begin{figure}
\figurenum{3}
\includegraphics[width=3.5in]{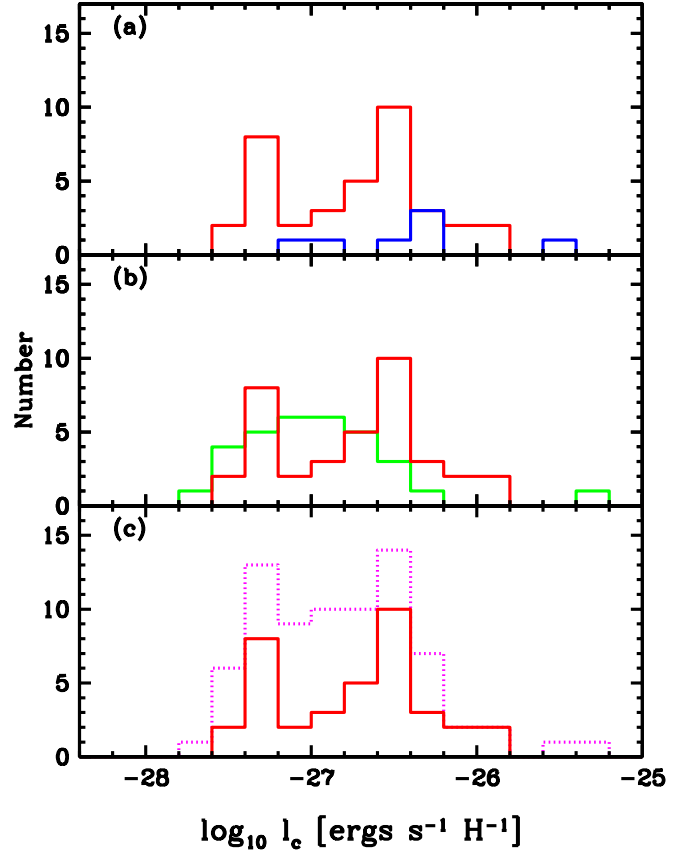}
\caption[]{Histograms comparing   
positive detections (red) and (a) lower limits (blue),
(b) upper limits (green), and (c)
all the {\lc} values in
Table 1 (magenta).
} 
\label{fig:lchist_all}
\end{figure}

These results were obtained by excluding 
both the lower limits and upper limits on {\lc}. This does
not affect our conclusions provided 
the true values of {\lc}, i.e., {\lctrue}, 
are
drawn from the same parent population as the positive
detections, which we assume is a bimodal distribution. 
Note by definition the values of {\lctrue} equal the measured
values of {\lc} in the case of positive detections, are less than
the measured {\lc} for the upper limits, and are greater than the measured
{\lc} for the lower limits.
The distributions of lower limits and positive detections shown in 
Fig.~{\ref{fig:lchist_all}}a are clearly compatible.
However, because the values of {\lctrue} must exceed the corresponding
lower limits, it is possible 
that the {\lctrue}
are drawn from a separate population concentrated above
the peak at {\lc} $\approx$ 10$^{-26.6}$ {\lcunit}
characterizing the positive detections.
But there is no sign in any  of the other properties
of DLAs with lower limits that distinguishes them from the DLAs with
positive detections. This suggests that both
sets of DLAs are drawn from a common population.
In any case the small number of lower limits 
implies they have negligible impact on the case for bimodality or
on the location of either peak inferred from the positive detections.  

By contrast, 
Fig.~{\ref{fig:lchist_all}}b shows 
that the  {\lc} distribution of the upper limits
peaks at  
$\approx$ 10$^{-27.0}$ {\lcunit}, the same value of {\lc}
where 
the positive
detections exhibit a trough. 
The  difference between the distributions is potentially
important owing to the comparable sizes  of the two samples.
However, the distribution of upper limits shown in
Fig.~{\ref{fig:lchist_all}}b 
does not accurately represent  the corresponding
distribution of {\lctrue}. 
Because the values of {\lctrue} 
should be lower than the associated
upper limits, 
the resulting distribution of {\lctrue}
should peak at values lower than shown in
the figure. While  we do not know how much lower, our recent
(2006 Dec., 2007 April, and 2007 Sept.) HIRES observations provide a clue. 
We observed
13 DLAs with upper limits set previously
by data acquired with ESI.
Using the higher spectral resolution of HIRES ($R$=43,000 compared to 
$R$=9000 for ESI), we  detected four of
these while the other nine remained undetected.
The HIRES observations moved five of 
eight ESI upper limits out of the trough to upper
limits below the trough, while two
of the four new detections were also moved below
the trough and one remained within the trough.
Therefore, it is plausible to assume that
{\lctrue} could typically be more than 0.4 dex
lower than the upper limits;  
i.e., the evidence is
consistent with a distribution resembling the positive
detections by exhibiting a peak at
{\lc}$\approx$10$^{-27.4}$ {\lcunit}.

Fig.~{\ref{fig:lchist_all}}c  plots the empirical {\lc} distribution
for all 76 DLAs in Table 1: the plot includes
the positive detections, lower limits, and upper limits.
While the corresponding {\lctrue} distribution  
for the entire sample has not fully been determined, the
above arguments provide good evidence for a `high cool'
peak at {\lc}$\approx$10$^{-26.6}$ {\lcunit}
and growing evidence for a `low cool' peak at {\lc}$\approx$10$^{-27.4}$
{\lcunit}; i.e., a distribution resembling that 
of the positive detections.
Although we do not know which values of {\lctrue}
to assign to the upper limits, we
suggest that in most cases the range of values is given by
{\lctrue}=[$10^{-28.5},10^{-27.0}$]
{\lcunit}. Lower bounds on {\lctrue} are set by the
heating rates due to FUV and X-ray background radiation
and excitation rates due to CMB radiation. The CMB 
sets a floor of {\lctrue}$\gtrsim$10$^{-28}$ {\lcunit}
for DLAs with $z$ $\gtrsim$3.5 (Wolfe {\etal} 2004;
hereafter referred to as WHGPL), whereas X-ray
heating provides the lower bound for DLAs with lower redshifts.
We reject values of {\lc} lower than
10$^{-28.5}$ {\lcunit} since the densities implied in 
the case of thermal balance 
$n$ $<$ 10$^{-1.5}$
cm$^{-3}$ (WHGPL).  
In that case
the length  
scale $d$ of gas clouds for DLAs with median
column density {\nh}=8{$\times$}10$^{20}$ cm$^{-2}$ 
(WGP05), would exceed 8 kpc. While the line-of-sight
might traverse such distances through DLAs, the multi-component
structure of the absorption-line  profiles indicates
the gas is confined to several smaller ``clouds'' characterized
by a low volume filling factor (Nagamine {\etal} 2007).
This scenario is supported by multi-phase models in which the
DLA gas density $n$ $>$ 10 cm$^{-3}$ and $d$ $<$ 30 pc
(Wolfe, Gawiser, Prochaska 2003; hereafter referred to
as WPG03). At the same time it is likely that {\lctrue} corresponding
to most of the upper limits in 
Fig.~{\ref{fig:lchist_all}}c is less than 10$^{-27}$ {\lcunit}:
since only four out of 32 upper limits exceeds 10$^{-26.6}$
{\lcunit} and we expect {\lctrue} to be more than 0.4 dex lower
than the corresponding upper limit, then {\lctrue} for
over $\approx$ 90$\%$ of the upper limits should be lower than 10$^{-27}$
{\lcunit}.

To summarize, standard statistical tests
applied to the {\lc} distribution of positive
detections support the bimodal hypothesis. The above
arguments concerning the lower and upper limits
are consistent with the hypothesis of two
peaks in the {\lctrue} distribution.
Because the presence of two peaks 
is supported
by physical arguments concerning
the range of {\lctrue},
the case
for bimodality is sufficiently compelling to
consider independent tests to which we now turn.

\section{INDEPENDENT TESTS FOR BIMODALITY}

If the {\lc} distribution is bimodal, the critical cooling
rate {\lccrit} should divide the DLA sample into independent
populations with physically distinct properties. In this
section we  investigate whether this is the case.

\begin{figure}[ht]
\figurenum{4}
\includegraphics[width=3.5in]{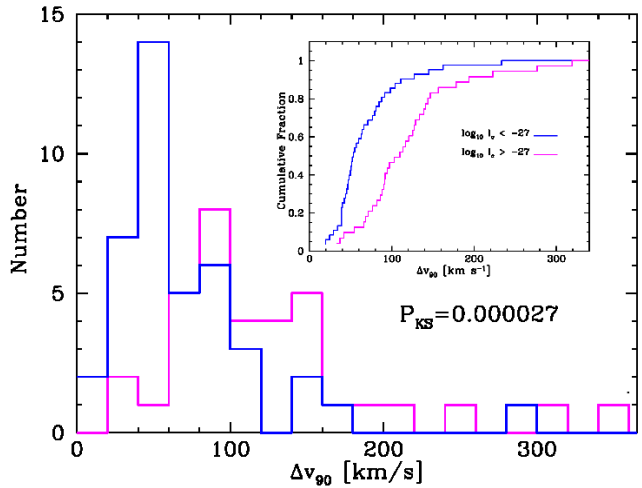}
\caption[]{Histograms  of velocity intervals {\dv} divided such
that the magenta histogram depicts DLAs with {\lc} $>$ {\lccrit} and
the blue histogram depicts DLAs with {\lc} $\le$ {\lccrit}. The inset
shows the cumulative distributions for both subsamples.}
\label{fig:delvdist}
\end{figure}

\subsection{Velocity Interval}

We first determine whether 
{\lccrit} divides DLAs into populations with distinct velocity
structures. 
We choose 
the absorption-line velocity interval {\dv} as a measure of
velocity structure, where {\dv} is 
defined as the velocity interval containing the central
90$\%$ of the optical depth of unsaturated low-ion absorption
lines (Prochaska \& Wolfe 1997): this definition guarantees that
{\dv} corresponds to the bulk of the neutral 
gas producing the absorption.
The measured values of {\dv} are listed in column 8 of Table 1.
To generate suitable subsamples we adopted the
following criteria. For the positive detections we assigned DLAs with
{\lc} $\le$ {\lccrit} to a `low cool' subsample and DLAs with
{\lc} $>$ {\lccrit} to a `high cool' subsample. 
The division is not as
clear for DLAs with upper limits because {\lctrue} may be
lower than {\lccrit} for objects with upper limits on {\lc}
exceeding {\lccrit}. To account for this possibility we assigned
DLAs with upper limits given by
log$_{10}${\lc} $\le$ log$_{10}${\lccrit} +{\delc}
to the `low cool' subsample
and those with higher upper limits to the `high cool' subsample.
Similarly we assigned DLAs with lower limits given by 
log$_{10}${\lc} $\le$ log$_{10}${\lccrit}$-${\delc} to the `low cool'
sample and those with higher lower limits to the
`high cool' subsample. The value of {\delc} is
determined by the ratio of the {\lc} limit
to our estimate of {\lctrue}.

\begin{table}[ht] \footnotesize 
\begin{center}
\caption{{\sc KS probabilities}}
\begin{tabular}{lcccccccc}
\tableline
\tableline
&&&\multicolumn{5}{c}{log$_{10}$($P_{\rm KS}$)$^{a}$}&\\
\cline{5-9}
log({\lccrit})$^{b}$ &{\delc} \ $^{c}$&$n_{lo}$$^{d}$ &$n_{hi}$$^{e}$& {\dv}  &[M/H] & {$\kappa$} &$W_{1526}$&  {\nh}  \\
\tableline
$-$27.1&0.0&28&48&$-$4.01&$-$3.19&$-$1.85&$-$2.36&$-$0.73\cr
$-$27.1&0.2&34&42&$-$3.71&$-$4.36&$-$3.40&$-$2.59&$-$0.69\cr
$-$27.1&0.4&39&37&$-$5.07&$-$5.45&$-$4.43&$-$2.81&$-$0.22\cr
$-$27.0&0.0&34&42&$-$3.92&$-$3.93&$-$1.91&$-$2.59&$-$0.69\cr
$-$27.0&0.2&40&36&$-$4.57&$-$4.34&$-$2.99&$-$3.14&$-$0.06\cr
$-$27.0&0.4&43&33&$-$4.35&$-$4.66&$-$3.30&$-$3.14&$-$0.02\cr
$-$26.9&0.0&39&37&$-$2.82&$-$2.75&$-$2.31&$-$2.23&$-$0.55\cr
$-$26.9&0.2&44&32&$-$4.35&$-$3.71&$-$3.30&$-$2.51&$-$0.07\cr
$-$26.9&0.4&47&29&$-$4.33&$-$4.05&$-$3.69&$-$2.84&$-$0.03\cr
$-$26.8&0.0&46&30&$-$4.24&$-$2.78&$-$2.42&$-$2.25&$-$0.75\cr
$-$26.8&0.2&49&27&$-$4.11&$-$3.11&$-$2.78&$-$2.25&$-$0.91\cr
$-$26.8&0.4&51&25&$-$4.39&$-$3.08&$-$2.80&$-$2.19&$-$0.02\cr
$-$26.7&0.0&49&27&$-$4.30&$-$3.11&$-$2.97&$-$2.25&$-$0.47\cr
$-$26.7&0.2&52&24&$-$4.04&$-$3.53&$-$3.31&$-$2.61&$-$0.07\cr
$-$26.7&0.4&53&23&$-$3.70&$-$3.07&$-$2.91&$-$2.19&$-$0.02\cr
\tableline
\end{tabular}
\end{center}
\tablenotetext{a}{KS probabilities  that parameters in `low cool' and `high cool' sub-sample
are drawn from same parent population}
\tablenotetext{b}{Critical cooling rate separating positive detections
in `low cool' and `high cool' subsamples}
\tablenotetext{c}{Correction to critical cooling rate for
upper limits (see text)}
\tablenotetext{d}{Number of DLAs in `low cool' sub-sample}
\tablenotetext{e}{Number of DLAs in `high cool' sub-sample}
\end{table}

Histograms with \delc$=0.2$ and {\lccrit}=10$^{-27.0}$ {\lcunit}
are shown in Fig.~{\ref{fig:delvdist}}. 
In this case the `low cool'  and `high cool' 
histograms are clearly different. Whereas the bulk
of the `low cool' subsample clusters around {\dv}$\approx$ 50 {\kms},
the bulk of the  `high cool' subsample DLAs exhibits a wider distribution
centered near 100 {\kms}. The difference between the two
subsamples is also
evident in the cumulative distribution
shown in the inset in 
Fig.~{\ref{fig:delvdist}}. The cumulative distributions are
obviously different with medians {\dv} = 46 {\kms}
for the `low cool' subsample and 104 {\kms} for the
`high cool' subsample. Applying the standard Kolmogorov-Smirnov (KS)
test we find the probability that the two distributions
are drawn from the same parent population, $P_{\rm KS}$({\dv})
=2.7{$\times$}10$^{-5}$; i.e., the null hypothesis
can be rejected at a high confidence level.

To test the sensitivity of this conclusion to uncertainties in
{\lccrit} and {\delc} we recomputed $P_{\rm KS}$({\dv}) for  a
range of values compatible with the {\lc} distribution in
Fig.~{\ref{fig:lchist_pd}}; i.e., 10$^{-27.1}$ $\le$ {\lccrit}
$\le$ 10$^{-26.7}$ {\lcunit} and 0 $\le$ {\delc} $\le$ 0.4.
The results shown in Table 2 indicate that 8.5{$\times$}10$^{-6}$
$\le$ $P_{\rm KS}$({\dv}) $\le$ 1.5{$\times$}10$^{-3}$, with the largest
value corresponding to the case {\delc}=0, which
is  unlikely. Next
we checked whether the {\dv} distribution was 
bimodal 
with respect to
other parameters measured in DLAs. The only
parameter suitable for this purpose
is redshift. Accordingly we divided our sample
into low-$z$ and high-$z$ subsamples
around the median redshift $z_{med}$=2.85. When we compared
the resulting {\dv} distributions, we found that
$P_{\rm KS}$({\dv}) = 0.69. In other words velocity
width is unlikely to be bimodal with respect
to redshift.

\begin{figure}[ht]
\figurenum{5}
\includegraphics[height=3.5in,angle=-90]{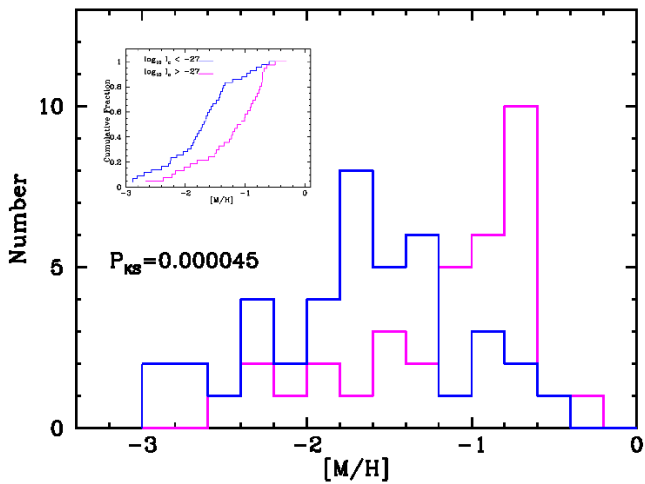}
\caption[]{Histograms  of DLA metallicity  divided such
that the magenta histogram depicts DLAs with {\lc} $>$ {\lccrit} and
the blue histogram depicts DLAs with {\lc} ${\le}$ {\lccrit}. The inset
shows the cumulative distributions for both subsamples.}
\label{fig:metaldist}
\end{figure}

\subsection{Metallicity}

Fig.~{\ref{fig:metaldist}} shows the metallicity distributions
of the `low cool' and `high cool' subsamples. The logarithmic
metal abundances with respect to solar, [M/H], are listed
in column 9 of Table 1, where M stands for an element
found to be undepleted in DLAs such as S, Si, or Zn 
(Prochaska {\etal} 2003).
As in  
Fig.~{\ref{fig:delvdist}} we divide the sample 
with {\lccrit}=10$^{-27}$ {\lcunit} and {\delc} =0.2.
Similar to the case of the {\dv} distributions
we find the `low cool'
and `high cool' distributions of metallicity are different.
For these values of {\lccrit}
and {\delc} the median metallicity of the `low cool' subsample
is given by [M/H]$_{low}=-1.74$, and [M/H]$_{hi}=-1.06$
for the `high cool' subsample.
Applying the KS test, we find the probability that the
two subsamples are drawn from the same parent population is
given by $P_{\rm KS}$([M/H])=4.5{$\times$}10$^{-5}$.
When we recomputed $P_{\rm KS}$([M/H]) for the range of
{\lccrit} and {\delc} shown in Table 2, we found that
3.5{$\times$}10$^{-6}$ $\le$ $P_{\rm KS}$([M/H]) $\le$ 1.8{$\times$}10$^{-3}$, 
where the largest value again corresponds to the unlikely
case of {\delc}=0. By contrast, when we split the sample according
to redshift we found that $P_{\rm KS}$([M/H])=0.50,
indicating that DLA metallicities in the two redshift
bins are not drawn from the same parent population.
This may conflict with our detection of 
metallicity evolution with redshift (Prochaska {\etal} 2003)
in which we found weak but statistically
significant redshift evolution; i.e.,
$d[{\rm M/H}]/dz$ = $-0.26$$\pm$0.07. Part of the difference between
the results could be due to sample size, since for the
Prochaska {\etal} (2003) sample, which is about twice the
size of the current sample,
we find
$P_{\rm KS}$([M/H])=0.001.
While it is possible 
that some of the more recent measurements may dilute
the original result, the vastly different
values of  
$P_{\rm KS}$([M/H]) is puzzling.

\begin{figure}[ht]
\figurenum{6}
\includegraphics[width=3.5in]{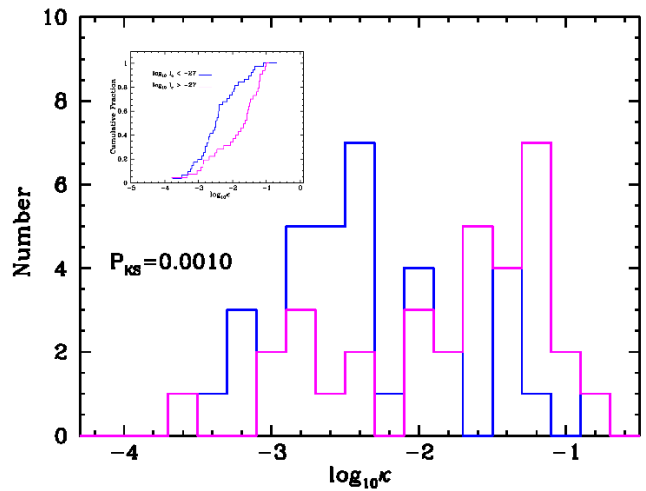}
\caption[]{Histograms  of DLA dust-to-gas ratio divided such
that the magenta histogram depicts DLAs with {\lc} $>$ {\lccrit} and
the blue histogram depicts DLAs with {\lc} ${\le}$ {\lccrit}. The inset
shows the cumulative distributions for both subsamples.}
\label{fig:kappadist}
\end{figure}

\subsection{Dust-to-Gas Ratio}

Not surprisingly the results for
dust-to-gas ratio, $\kappa$, resemble
the results for metallicity.
This is shown in
Fig.~{\ref{fig:kappadist}}, which compares the
$\kappa$ distributions for the `high cool' and `low cool'
subsamples where 
{$\kappa$}=
10$^{\rm [M/H]}$(10$^{\rm [Fe/M]_{\rm int}}$-10$^{\rm [Fe/M]}$)
(see WPG03) and 
[Fe/M] and [Fe/M]$_{\rm int}$ are the measured
and intrinsic logarithmic abundances of
Fe with respect to an undepleted $\alpha$ element, M:
all abundances
are with respect to solar ([Fe/H] is given in column 10 in
Table 1). 
We computed the numerical
values for $\kappa$ in Table 2 
by assuming
[Fe/M]$_{\rm int}$ 
=$-$0.2 when [Fe/M] $\le  -0.2$ and
[Fe/M]$_{\rm int}$ =0 when
$-$0.2 $\le$ [Fe/M] $\le$ 0.0 (see WGP05). 

Fig.~{\ref{fig:kappadist}} shows that the dust-to-gas
ratios of the `high cool' subsample are significantly higher
than for the `low cool' subsample. 
We find $P_{\rm KS}(\kappa)$=
1.0{$\times$}10$^{-3}$ 
for the
distributions shown in 
Fig.~{\ref{fig:kappadist}}, where 
{\lccrit}=10$^{-27.0}$ {\lcunit}
and {\delc}=0.2. 
We also find that
 3.7{$\times$}10$^{-5}$ $<$ $P_{\rm KS}(\kappa)$ $<$  1.4{$\times$}10$^{-2}$
for the range in {\lccrit} and {\delc} in Table 2,
with the largest values again given by the unlikely case
{\delc}=0. 
When
we divided the sample according to redshift we found
that $P_{\rm KS}(\kappa)$=0.17. 
Although a
unimodal distribution of $\kappa$ with
respect to redshift is less likely, it still cannot be
ruled out with high significance. 
As a result bimodality is unlikely to arise with respect to redshift. 

\begin{figure}[ht]
\figurenum{7}
\includegraphics[width=3.5in]{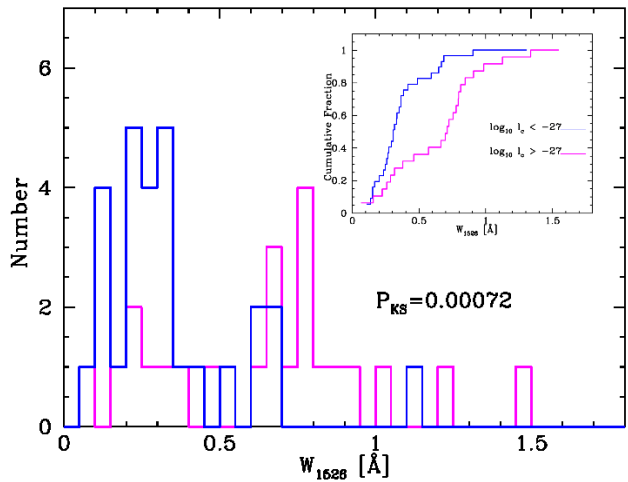}
\caption[]{Histograms  of Si II $\lambda$1526 equivalent width 
divided such
that the magenta histogram depicts DLAs with {\lc} $>$ {\lccrit} and
the blue histogram depicts DLAs with {\lc} ${\le}$ {\lccrit}. The inset
shows the cumulative distributions for both subsamples.}
\label{fig:W1526}
\end{figure}

\subsection{Si II $\lambda$ 1526 Equivalent Width}

Prochaska {\etal} (2007) recently showed that DLAs exhibit
a tight correlation between Si II $\lambda$ 1526 rest-frame
equivalent width, $W_{1526}$, and metallicity, [M/H]. Interestingly
the correlation exhibits less scatter than the {\dv}
versus [M/H] correlation. This is a striking result because 
whereas {\dv} and [M/H] are  determined from the same neutral gas,
$W_{1526}$ is dominated by the kinematics of low optical-depth
clouds that make an insignificant contribution to [M/H].
Prochaska {\etal} (2007) argue that $W_{1526}$ is determined
by the virialized random motions of clouds in the outer 
halo of the DLA galaxy, while {\dv} is set 
by the motions (e.g.\ rotation)
of the neutral ISM. As a result, the wide range of values expected
for impact parameter and galaxy inclination will cause sizable
scatter in {\dv} among galaxies with the same virial velocity. By
contrast, the scatter in $W_{1526}$ will be lower for sightlines
encountering large numbers of randomly moving clouds in
the halos of the same galaxies. Because of the tight
correlation between $W_{1526}$ and [M/H],
one might expect that $W_{1526}$ determined for the
'low cool' and `high cool' DLAs are not drawn from
the same parent population.  

Fig.~{\ref{fig:W1526}} 
compares the {\w1526} distributions
of the `high cool' and `low cool' subsamples for the standard
values {\lccrit}=10$^{-27}$  {\lcunit} and {\delc}=0.2. In this
case 
the 51 measured values of {\w1526} are listed in column 11 of Table 1. 
The figure shows that the `low cool' subsample exhibits systematically lower
values of {\w1526} than the `high cool' subsample: we find
$P_{\rm KS}$({\w1526})=7.2{$\times$}10$^{-4}$. When we
recomputed 
$P_{\rm KS}$({\w1526}) for the range of {\lccrit} and
{\delc} shown in Table 2, we found that 7.2{$\times$}10$^{-4}$ $<$
$P_{\rm KS}$({\w1526}) $<$ 6.5{$\times$}10$^{-3}$. When we split
the sample according to redshift we found $P_{\rm KS}$({\w1526})
=0.87, which further supports the hypothesis that the
DLA sample is not bimodal with respect to redshift. 

\begin{figure}[ht]
\figurenum{8}
\includegraphics[width=3.5in]{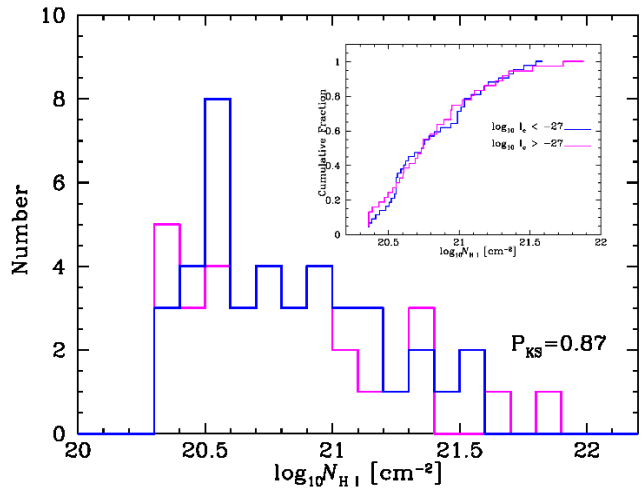}
\caption[]{Histograms  of H I 
column density  divided such
that the magenta histogram depicts DLAs with {\lc} $>$ {\lccrit} and
the blue histogram depicts DLAs with {\lc} ${\le}$ {\lccrit}. The inset
shows the cumulative distributions for both subsamples.}
\label{fig:NHIdist}
\end{figure}

\subsection{H I Column Density}

Fig.~{\ref{fig:NHIdist}}, compares the {\nh} distributions
of the `high cool' and `low cool' subsamples for the standard
case {\lc}=10$^{-27}$ {\lcunit} and {\delc}=0.2,
where the values of {\nh} are listed in column 5 of Table 1.
By contrast with the previous results, the null
hypothesis cannot be excluded at high confidence levels. 
Specifically, in this
case $P_{\rm KS}$({\nh})=0.87. More generally,
0.19 $<$ 
$P_{\rm KS}$({\nh}) $<$ 0.96 for the range of values
spanned by {\lccrit} and {\delc} in Table 2.
This has important implications to be discussed
in $\S$ 5.
When we split the data according to redshift we found
that $P_{\rm KS}$({\nh})=0.21, indicating that we cannot
confidently rule out that {\nh} values in the two
redshift bins are drawn from the same parent population.
As a result, the DLA sample is unlikely to be bimodal
with respect to redshift.

\section{MODES OF HEATING }

In this section we describe the heating processes 
that balance cooling
in DLAs. We interpret the physical
significance of the critical cooling rate
{\lccrit} and discuss differences
between the heat
input into the `low cool' and `high cool' DLAs. 
We consider two possibilities.

\subsection{Background Heating of `Low Cool' DLAs}
 
The first possible explanation for the presence of a
trough at {\lccrit} $\approx$ 10$^{-27}$ {\lcunit} 
is that {\lccrit} is the maximum [C II] 158 {\micron}
emission rate of low-metallicity gas heated by background
radiation at $z$ $\sim$ 3. According to
 this scenario, DLAs with
{\lc} {$\le$} {\lccrit}, i.e., the `low cool' DLAs, are
neutral gas layers heated
by background radiation alone, while DLAs with
{\lc} $>$ {\lccrit}, the
`high cool' DLAs, 
are in addition  heated by internal sources
(see WHGPL and Wolfe \& Chen 2006; hereafter referred to as
WC06).

\begin{figure}
\figurenum{9}
\includegraphics[width=3.5in]{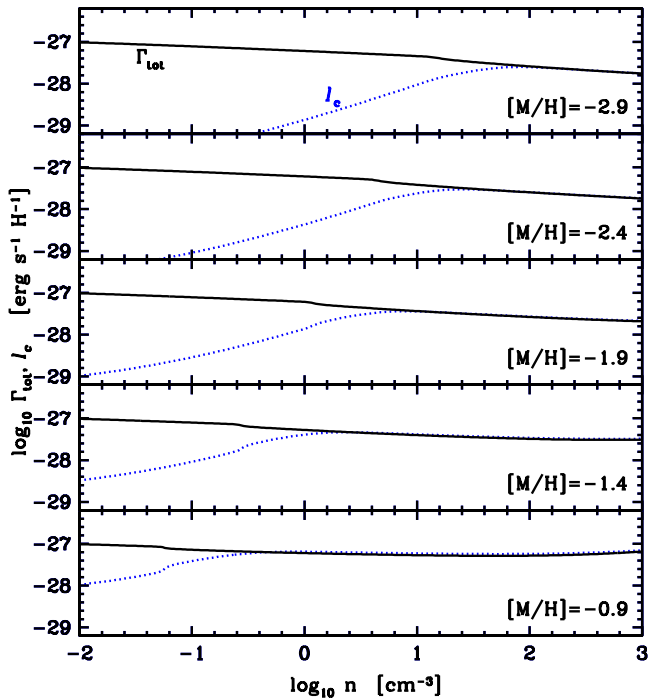}
\caption[]{Thermal equilibria computed for neutral gas
at $z$ $\sim$ 3 heated by background radiation. Dominant
heating mechanism is production of
primary electrons by photoionization of H and He by soft
X-rays ($h{\nu} \ge $ 0.4 kev). 
Primary electron energy degraded by
photoionization and collisional excitation processes, leading
to the production of secondary electrons that heat ambient
electrons by Coulomb interactions. 
Grain photoelectric
heating by FUV radiation is also included. 
Background radiation fields computed
by Haardt \& Madau (2003; using CUBA). Panels show the background
heating rates per H atom  (solid black)
and the [158] {\micron} cooling rates, {\lc}, per H atom (dotted blue)
as functions of gas  density. 
Panels show results for several values
of gas metallicity, [M/H]. Notice, {\lc}
does not exceed 10$^{-27}$ {\lcunit}.
See text for physical interpration
of these curves.} 
\label{fig:critcooling}
\end{figure}

This interpretation of {\lccrit} is illustrated in 
Fig.~{\ref{fig:critcooling}}, which  
shows {\lc}  
and the total heating rate, $\Gamma_{\rm tot}$,
as functions of gas volume density $n$.
We assume that
$\Gamma_{\rm tot}$=$\Gamma_{\rm XR}$$+$
$\Gamma_{\rm pe}$$+$$\Gamma_{\rm C^{0}}$, where
$\Gamma_{\rm XR}$, 
$\Gamma_{\rm pe}$,
and $\Gamma_{\rm C^{0}}$
are heating rates due to X-ray photoionization, grain
photoelectric emission, and 
photoionization
of C$^{0}$  (Wolfire {\etal} 1995). The curves in
Fig.~{\ref{fig:critcooling}}  
are solutions to
the thermal balance and ionization
equilibrium equations for neutral gas layers with
{\nh}=10$^{20.6}$ cm$^{-2}$ that are
heated and ionized by background radiation fields
computed for $z$=2.3 (Haardt \& Madau 1996: and more recently
using CUBA\footnote{CUBA(Haardt \& Madau 2003) is available
at:
http://pitto.mib.infn.it/$\sim$haardt/cosmology.html}).
The panels in this figure show the resulting
$\Gamma_{\rm tot}(n)$ and {\lc}($n$) for metallicities
spanning the range of values found in DLAs.

Fig.~{\ref{fig:critcooling}} 
illustrates why background heating
is an attractive explanation for DLAs with {\lc} $\le$
{\lccrit}.  In particular  
{\lc} remains below
10$^{-27}$ {\lcunit} for the entire range of densities
and metallicities in the figure. At densities below
$n_{\rm max}$, 
the lowest density  where {\lc} is  a local maximum (={\lcmax}), 
the gas is a warm
neutral medium (WNM) in which $T$ $\approx$ 8000 K and
{\lc} $<<$
$\Gamma_{\rm tot}$, since cooling is dominated by {\lya}
and electron recombination onto grains.
At densities above $n_{\rm max}$ the gas is a cold
neutral medium (CNM) in which $T$ $\approx$ 100 K and
{\lc}=
$\Gamma_{\rm tot}$, because cooling is dominated by 
158 {\micron} emission. While a detailed discussion about the
behavior of {\lc}($n$) is beyond the scope of this paper
(see WHGPL and Wolfire {\etal} [1995]  for more extensive discussions),
we wish to emphasize two points.
First, although {\lcmax} increases with increasing [M/H],
Fig.~{\ref{fig:critcooling}} shows that
it never exceeds
10$^{-27}$ {\lcunit}. Second, because {\lc} undergoes a sharp
decline with {\it decreasing} density at $n$ $<$ $n_{\rm max}$,
the range of 158 {\micron} cooling rates available in the
WNM for
low values of {\lc} is much larger than the limited range
of CNM cooling rates available at
$n$ $>$ $n_{\rm max}$. Combined with the upper limits
on density  required
in the case of background heating to explain the
large observed ratios of C II/C I (WHGPL),
this suggests that DLAs with {\lc}
$\le$ {\lccrit} are low-density WNM gas layers without local
heat input. 

Because these conclusions are based on specific values
of {\nh} and $z$, we tested their generality by computing
{\lcmax}
for the 40 DLAs in the standard `low cool' population (see $\S$ 2).
We adopted the metallicity, dust-to-gas
ratio, and {\nh} appropriate for each DLA,  exposed the
gas to the 
background radiation intensity  
inferred
for the DLA redshift (Haardt \& Madau 2003; using CUBA), and then
computed {\lcmax} for the {\lc}($n$) equilibrium curves. 
We found that  
{\lcmax} exceeded 10$^{-27}$ {\lcunit}
for only 5 of the 40 DLAs.
This is consistent with our expectation,
owing to the  low metallicity of the `low cool' population.

However, despite the attractive features of
this model it has serious problems. 
While it predicts {\lc} to be
less than {\lcmax}, 
the {\it measured}
values of {\lc} exceed {\lcmax} for 9 out of the 14 positive
detections in the `low cool' population. This indicates
the presence of local heat sources for these DLAs,
which contradicts the assumption of background heating alone. 
Of course this conclusion is based on our determinations
of {\lcmax} which for low-metallicity gas is a sensitive function of the X-ray
background intensity at $z$$\approx$3 and photon energies
$h{\nu}$ $\ge$ 0.5 keV.
Because the background intensity  depends on volume emissivity, the
critical quantity is the normalization
of the X-ray luminosity function  for high-$z$
AGNs at the characteristic luminosity, $L_{*}$.
Results from the
recent study of Silverman {\etal} (2006; 2007) suggest that
the Haardt-Madau (2003; using CUBA) determination, which was based
on earlier work by
Ueda {\etal} (2003), is a factor of 3 higher than the
Silverman {\etal} (2006; 2007) determination for 3.0 $<$ $z$ $<$ 4.0,
the relevant redshift interval for background radiation
at $z$ $\approx$ 3. 
In addition the Haardt-Madau (2003; using CUBA) background must be increased
by at least a factor of three for background heating to explain
the measured values of {\lc} for the `low cool' population. 
As a result, the X-ray luminosity function must be a factor of
9
higher than the Silverman {\etal} (2006;  2007) value to explain the
{\lc} values.
While
there are other uncertainties in determining the heating rates
such as the carbon abundance and the grain-photoelectric
heating efficiencies, we conclude that while it cannot
be ruled out altogether, the
background heating hypothesis is not likely to be correct.

\subsection{ Local Heating}

We next consider the alternative hypothesis 
that the `low cool' DLAs
are  heated primarily by local sources.  
Suppose the background intensity, {\jnubkd}, at $z$ $\approx$ 3 is
half the Haardt \& Madau (2003; using CUBA) intensity,
in agreement with observational uncertainties.
In that case the measured values of {\lc}
exceed {\lcmax} for 11 of the 14 positive detections,
indicating that local energy input is required for
most of this population.
To determine the level of input we adopted the
model of WPG03 in which the star formation rate per unit
area projected  perpendicular to a uniform gaseous disk, 
{\sigmasfr} \footnote{In WC06 we denoted this quantity
by ({\ps})$_{\perp}$ to distinguish it from {\ps}, the SFR per unit
area determined from the observed H I column density by
assuming the Kennicutt-Schmidt relation. Since such distinctions
are not relevant here and to be consistent with the
more general usage (cf Kennicutt 1998) we adopt the {\sigmasfr} notation.},
generates FUV radiation that heats the gas
by the grain photoelectric mechanism.
Heat inputs by cosmic rays and locally
generated X-rays are also included and are
assumed to be proportional to {\sigmasfr}.
The gas is a two phase medium in which
the CNM and WNM are in pressure equilibrium
at the pressure given by   the geometric
mean of the minimum and maximum pressures
characterizing the equilibrium $P(n)$ curve
(Wolfire {\etal} 2003).
Because 
the detected {\ciis} absorption likely arises
in the CNM,
WPG03 were able to deduce unique values
for the local FUV radiation intensity
{\jnulocal} 
from measured values of {\lc}. Specifically, in the
case of thermal balance WPG03 found that 

\begin{equation}
{\ell_{c}}=10^{-5}{\kappa}{\epsilon}J_{\nu} 
\end{equation}

\noindent where 
{\jnu}={\jnubkd}$+${\jnulocal} and $\epsilon$ is the photoelectric
heating efficiency 
(e.g. Bakes \& Tielens 1994; Weingartner \& Draine 2001). To obtain
{\sigmasfr}, WPG03 solved the transfer equation in
a uniform disk and found that in the optically thin limit

\begin{equation}
J_{\nu}^{\rm local}={C{\Sigma_{\rm SFR}} \over {8{\pi}}}[1+{\ln}(r_{\rm DLA}/h)]
\end{equation}

\noindent where
$C$ ($\equiv$8.4{$\times$}10$^{-16}$
ergs cm$^{-2}$s$^{-1}$Hz$^{-1}$[{\msolar}yr$^{-1}$kpc$^{-2}$]$^{-1}$)
is a conversion constant (see WC06),
and $r_{\rm DLA}$ and $h$ are the radius and scale-height of
the DLA disk.

\begin{figure}[ht]
\figurenum{10}
\includegraphics[width=3.5in]{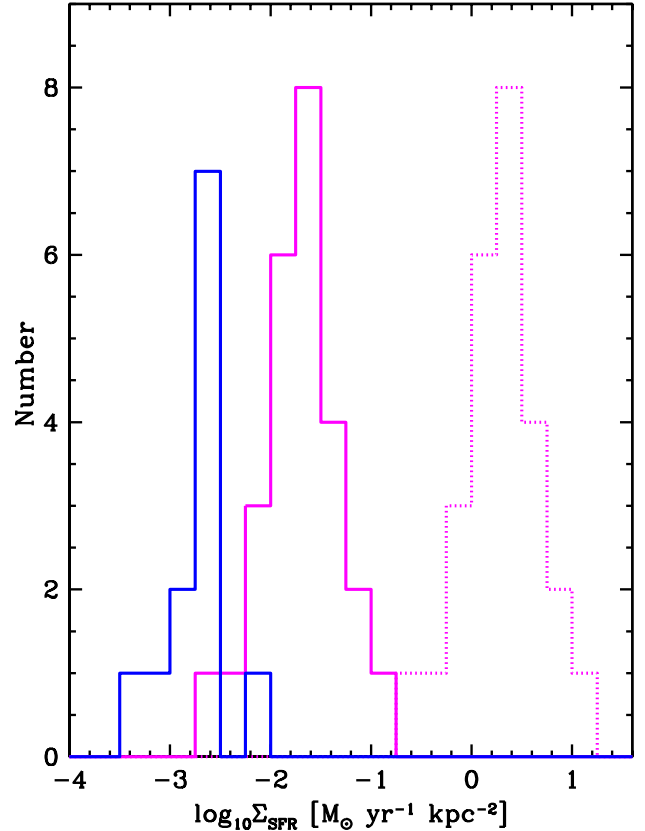}
\caption[]{Histograms  of {\sigmasfr} obtained
using the {\ciis} model of WP03 and WGP03. Blue histogram
refers to `low cool' DLAs. Magenta histograms refer to `high cool'
DLAs, where solid and dotted histograms respectively depict lower
limits and  true values of
of {\sigmasfr}.} 
\label{fig:SFRhist}
\end{figure}

Applying the same technique to the DLA sample in Table 1
and assuming $r_{\rm DLA}/h$=50,
we determined values of {\sigmasfr} for 38 of the 44 DLAs with positive 
detections and lower limits. \footnote {We could not determine
{\sigmasfr} for the remaining 6 DLAs either because   background 
heating was sufficient to
balance the observed cooling, or we could not deduce a
dust-to-gas ratio from our depletion model since
[Fe/H] $>$ [M/H].}
The resulting distribution of {\sigmasfr} is
shown in 
Fig.~{\ref{fig:SFRhist}}, where we have plotted the
results for the `low cool' and `high cool' populations separately. 
We include results for the `high cool' DLAs,
since background heating is insufficient to balance the 158 {\micron}
cooling rates from this population, indicating that
local heating is also required in this case.
The {\sigmasfr} distribution is of interest for several 
reasons. First,  application of the KS 
test  provides strong evidence for two distinct modes
of star formation, since the probability that the
two populations are drawn from the same parent
distribution is given by
$P_{\rm KS}$({\sigmasfr})=3.4{$\times$}10$^{-7}$. Furthermore, comparison of
Fig.~{\ref{fig:SFRhist}} with 
Figs. 3 to 6
shows evidence for less overlap
between the `low cool' and `high cool' distributions of
{\sigmasfr} than for {\dv}, [M/H], $\kappa$, and $W_{\rm 1526}$.
Indeed, comparison with Fig. 2 suggests
that 
the {\sigmasfr} and {\lc} distributions are both bimodal.

Second, although the WPG03 model assumes
{\it in situ} star formation throughout the neutral-gas
disk, the survey of the Hubble Ultra Deep Field (UDF) by WC06 places 
upper limits on  such star formation, which
rule out some of the scenarios discussed above.
Specifically, WC06 searched the UDF F606W image for
low surface-brightness emission from DLAs in the redshift
interval $z$=[2.5,3.5] with angular (linear) diameters between
$\theta_{\rm DLA}$=0.25 {\arcsec} ($d_{\rm DLA}$=2 kpc)
and
$\theta_{\rm DLA}$=4.0 {\arcsec} ($d_{\rm DLA}$=31 kpc).
They optimized the survey sensitivity by convolving the F606W
image with Gaussian smoothing kernels with FWHM diameters
$\theta_{\rm kern}$=
$\theta_{\rm DLA}$: the resulting $\approx$5$\sigma$ surface brightness
thresholds varied between $\mu_{V}^{\rm thresh}$=28.0 mag arcsec$^{-2}$
for 
$\theta_{\rm kern}$=0.25\arcsec\ to
$\mu_{V}^{\rm thresh}$=29.7 mag arcsec$^{-2}$
for $\theta_{\rm kern}$=4.0\arcsec. They found the number of 
detected objects to decrease steeply with increasing
$\theta_{\rm kern}$ and to be at least two orders of magnitude
lower than predicted, which is determined
by dividing the large DLA area covering
factor, $C_{\rm A}$=0.33 (for $z$=[2.5,3.5]), by
the area per DLA, ${\pi}{\theta_{\rm DLA}}^{2}/4$. Stated differently,
{\it in situ} star formation is unlikely to occur in
DLAs at values of {\sigmasfr} above the SFR thresholds,
$\Sigma_{\rm SFR}^{\rm thresh}$, set
by the UDF threshold surface brightnesses. To compute 
$\Sigma_{\rm SFR}^{\rm thresh}$
we note that 
$I_{\nu_{0}}$=$C${\sigmasfr}[4${\pi}(1+z)^{3}$cos($i$)]$^{-1}$ where
$I_{\nu_{0}}$ is the intensity observed from a disk at redshift $z$,
and
$i$ is the disk inclination angle. Therefore, a disk
with a given  {$\Sigma_{\rm SFR}^{\rm thresh}$} generates a
range of
intensities that depend on $i$. To
assure the detection of a significant fraction of disks with  
threshold SFRs we set $i$=60$^{o}$. This guarantees that  $I_{\nu_{0}}$
$\ge$ $I_{\nu_{0}}^{\rm thresh}$ 
(or $\mu_{V}  \le \mu_{V}^{\rm thresh}$) for 
half the disks for  which {\sigmasfr}
={$\Sigma_{\rm SFR}^{\rm thresh}$} and $i \ {\ge}$ 60$^{o}$.

\begin{figure}[ht]
\figurenum{11}
\includegraphics[width=3.5in]{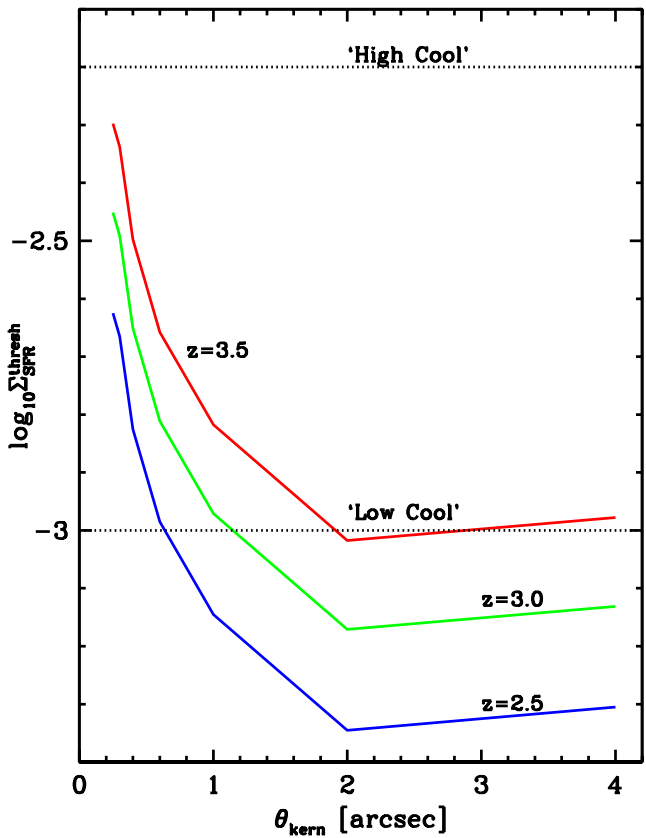}
\caption[]{Solid curves are
{\it threshold} values of {\sigmasfr} versus
diameter of smoothing kernel used by the search of
WC06. Results shown for 
$z$ = 2.5 (blue), 3.0 (green), and 3.5 (red). 
Horizontal dotted lines depict cooling rates per H
atom characterizing `low cool' and `high cool' populations.}
\label{fig:sigma_sfr}
\end{figure}

The resulting $\Sigma_{\rm SFR}^{\rm thresh}$ are plotted 
as functions of 
$\theta_{\rm kern}$ in
Fig.~{\ref{fig:sigma_sfr}} 
for three redshifts spanning the redshift interval of the
WC06 survey.\footnote{For a given
surface brightness the above definition results in values of  
$\Sigma_{\rm SFR}^{\rm thresh}$ that are a factor of 2.8
smaller than the effective SFR per unit area, {\ps}, discussed
in WC06 (see their eq. 3).} 
The figure shows that {$\Sigma_{\rm SFR}^{\rm thresh}$}
decreases in the interval $\theta_{\rm kern}$=[0.25{\arcsec},2.0{\arcsec}]. 
The decrease is caused by the 
Poisson error in surface brightness, which decreases 
with increasing aperture size. The decline
of 
$\Sigma_{\rm SFR}^{\rm thresh}$ ceases at {$\theta_{\rm kern}$
$>$ 2.0{\arcsec} due to the appearance of the systematic error, 
which WC06 attribute to confusion noise 
arising from the low surface-brightness outer regions of
bright galaxies.
Comparison with  
Fig.~{\ref{fig:SFRhist}} shows that 
the SFR per unit area for the bulk of the `high cool' population,
exceeds 10$^{-2.2}$ {\sfrunit}. Since this exceeds
$\Sigma_{\rm SFR}^{\rm thresh}$
for all values of the redshifts and $\theta_{\rm kern}$ depicted
in 
Fig.~{\ref{fig:sigma_sfr}}, such star formation would have been
detected by WC06. 
Because WC06 detected less than 1$\%$ of the
DLAs predicted for this population, we conclude that the heat input from
{\it in situ} star formation in `high cool' DLAs cannot
balance the rate at which they cool.
This is in agreement with the
results of WC06 who worked with the Kennicutt-Schmidt
relation rather than the {\ciis} technique. 
As a result, the FUV intensity that heats the gas is likely
emitted by compact (i.e., $\theta_{\rm FWHM}$ $<$ 0.25 {\arcsec}), 
centrally located, regions that
WC06 identified as Lyman Break Galaxies (hereafter referred to
as LBGs; Steidel {\etal} 2003). 
Note, in that case the
{\sigmasfr} obtained for this population 
(see Fig.~{\ref{fig:SFRhist}}) are lower limits on
the true values of {\sigmasfr}, which we estimate
from the following
argument. Assuming the `high cool' DLAs account for
half the DLA population and contain LBGs, the ratio
of areas occupied by  DLAs to LBGs, 
${\pi}r_{\rm DLA}^{2}/{\pi}r_{\rm LBG}^{2}$, is
given by the ratio of area covering factors,
$C^{\rm high cool}_{\rm DLA}/C_{\rm LBG}$. Since WC06 found that
$C_{\rm DLA}/C_{\rm LBG}$=330 in the redshift
interval $z$=[2.5,3.5], we find that
${\pi}r_{\rm DLA}^{2}/{\pi}r_{\rm LBG}^{2}$=165, 
where
$C^{\rm high cool}_{\rm DLA}$=
0.5$C_{\rm DLA}$.
As a result,
the true values of {\sigmasfr} are a  factor
of 165
higher than the lower limits for the 
`high cool' population.  
The distributions of both 
the lower limits and true values of {\sigmasfr} 
are depicted in
Fig.~{\ref{fig:SFRhist}}. 

Fig.~{\ref{fig:SFRhist}} also shows that 
$\Sigma_{\rm SFR}$
$\ge$ 10$^{-3}$ {\sfrunit} 
for the bulk of the `low cool' population.
From Fig.~{\ref{fig:sigma_sfr}} we see that
this exceeds  $\Sigma_{\rm SFR}^{\rm thresh}$
for $\theta_{\rm kern}$ $\ge$ 0.7{\arcsec} at $z$=2.5,
$\theta_{\rm kern}$ $\ge$ 1.3{\arcsec} at $z$=3.0, and
$\theta_{\rm kern}$ $\ge$ 2.0{\arcsec} at $z$=3.5. Therefore,
{\it in situ} star formation is detectable from `low cool'
DLAs with angular diameters $\theta_{\rm DLA}$ above
these limits. Since less than 1 $\%$ of the predicted number
of DLAs were detected by WC06 and since the `low cool'
population comprises about half of all high $z$ DLAs, we
conclude that {\it in situ} star formation for this population
is compatible with the observations only if the values
of $\theta_{\rm DLA}$ are smaller than these limits.
Note, we cannot rule out the possibility 
that similar to the `high cool' population, star
formation in `low cool' 
DLAs is sequestered away from the DLA gas in compact LBG cores.
However, owing to the different properties of the two
populations, it is more likely that 
within
the context of the local heating hypothesis
the `low
cool' DLAs undergo {\it in situ}
star formation.

\clearpage



\begin{table}[ht] \footnotesize 
\begin{center}
\caption{{\sc Population Properties}}
\begin{tabular}{lcc}
\tableline
\tableline
& \multicolumn{2}{c}{Population}\\
\cline{2-3}
Property(median) &`Low Cool'& `High Cool' \\
\tableline

log$_{10}${\lc} \ [{\lcunit}]&$-$27.29$\pm$0.07 &$-$26.54$\pm$0.13\cr
{\dv} \ [{\kms}]&46$\pm$10 &104$\pm$15\cr
[M/H]&$-$1.74$\pm$0.19&$-$1.06$\pm$0.13\cr
{\w1526} \ [{\AA}]&0.26$\pm$0.09&0.67$\pm$0.19\cr
log$_{10}${\nh} \ [cm$^{-2}$]&20.7$\pm$0.17 &20.7$\pm$0.19 \cr
\cline{1-3}
log$_{10}$$\kappa$&$-$2.57$\pm$0.17 &$-$1.60$\pm$0.26 \cr
log$_{10}$[{\jnulocal}/10$^{-19}$ (cgs)]&$-$18.58$\pm$0.17&$-$17.71$\pm$0.12\cr
log$_{10}${\sigmasfr}$^{a}$ \ [{\sfrunit}]&$-$2.78$\pm$0.34 &$>$$-$1.85 \cr

\tableline
\end{tabular}
\end{center}

\tablenotetext{a}{SFRs per unit area inferred assuming {\it in situ}
star formation. Since {\it in situ} star formation ruled out for
`high cool' DLAs, entry for `high cool' DLAs is a lower limit. }
\end{table}

\section{SUMMARY OF DLA PROPERTIES}

At this point we pause to give an overview of the DLA properties
discussed above. A summary is given in Table 3 which lists
the median values of the directly measurable properties {\lc}, 
{\dv}, [M/H], $W_{\rm 1526}$, and {\nh}, and the model-dependent
quantities $\kappa$, {\jnulocal}, and {\sigmasfr}.
With the exception of {\nh}, all the
properties of the `high cool' DLAs have systematically
larger values than their `low cool' counterparts.
Because {\dv} and {\w1526} measure
spatial variations in velocity, it is plausible to assume 
they measure mass. In that case the median halo mass of 
the `high cool' DLAs would be 30 times larger than
that of the `low cool' DLAs. This follows from the assumptions 
that (1) {\dv} and {\w1526} are each proportional to the halo circular
velocity $v_{\rm circ}$ and (2) $M_{\rm DM}$ $\propto$ $v_{\rm circ}^{3}$ 
(see Mo {\etal} 1998) where {\mdm} is the halo dark-matter
mass. A higher mass for the `high cool' population
is also consistent with its higher metallicity, since metallicity
is an increasing function of mass in modern galaxies (Tremonti
{\etal} 2004),
and  chemical evolution in $\Lambda$CDM models predict
a similar trend at high redshifts (e.g. Nagamine  2002). 
Because the dust-to-gas ratio $\kappa$ traces metallicity, the
higher median value of $\kappa$ for the `high cool' DLAs is
expected. On the other hand  higher values of
{\lc} and {\sigmasfr} in the `high cool' DLAs goes against
the trend in current galaxies where star-formation activity
is more common in low-mass late type galaxies than in
massive early-type galaxies (Kauffmann {\etal} 2003).

While we have interpreted the two types of DLAs as physically
distinct populations, let us consider the alternative hypothesis
that they are distinguished by the difference in impact
parameters of sightlines passing through a
unique population of DLAs. Suppose 
`low cool' DLAs are  metal-poor outer regions connected
to metal-rich inner regions of the `high cool' DLAs
by a negative metallicity gradient. In this case
the lower values of {\dv}
in the `low cool' DLAs
would arise from passage of QSO sightlines through
neutral gas at larger impact parameters where smaller changes
in velocity generated by the systematic
motions of the neutral gas are expected (Prochaska \& Wolfe 1997).
There are, however, at least two difficulties with this idea.
First, since the values of {\w1526} are plausibly determined by the virialized
random motions of clouds in the outer halo, {\w1526} would 
not be affected by a change
in impact parameter, since $r_{\rm DLA}$ is small compared to the radius
of the halo. 
Yet the median value of {\w1526}
for the `low cool' DLAs is   
is significantly lower than for the `high cool' DLAs (see Table 3).
Secondly, one would also
expect a decrease in {\nh}
to accompany the decrease in [M/H] with increasing impact parameter, 
which 
is contradicted by the good agreement 
between the median values
of {\nh} in Table 3 (but see Johansson \& Efstathiou 2003).
Third, it is difficult to understand how a difference in
impact parameter could produce bimodality in {\lc}.

Therefore, while we cannot rule out the alternative possibility
that 
the velocity fields in DLAs could be due
to non-gravitational motions (e.g.\ galactic-scale winds,
see discussion in $\S$ 7), 
we shall adopt the working hypothesis that the
crucial quantity distinguishing the two populations is mass.

\section{ CONNECTIONS TO BIMODALITY IN 
GALAXIES}

In this section we discuss analogies with bimodality
in modern galaxies, and then place the results within
the context of galaxy formation.

\subsection{Bivariate Distributions}

The bimodality of the {\lc} distribution in 
Fig.~{\ref{fig:lchist_pd}} brings to mind 
bimodal distributions in contemporary galaxies.
From their analysis of $\sim$10$^{5}$ galaxies drawn
from  the SDSS
survey, Baldry {\etal} (2004; see also Blanton {\etal} 2003) find the $u-r$ color distribution
to be bimodal. The distribution exhibits a blue
peak at
$u-r$ $\approx$ 1.4  
and  a red peak at 
$u-r$ $\approx$ 2.4  . In a similar analysis, 
Kauffmann {\etal} (2003) find bimodality
in the distribution of the age parameter $D_{n}$(4000) and the
indicator of recent star-formation activity, H${\delta}_{A}$.
The galaxies centered at the blue peak
are roughly the same objects exhibiting
peaks at young stellar ages and recent star-formation, while
the 
the galaxies centered at the red peak
exhibit peaks at
older stellar ages and low levels of recent star-formation
activity.

A possible link to bimodality in DLAs follows from the
bivariate galaxy distributions in the ($u-r$,$M_{*}$)
\footnote{Baldry {\etal} actually use the $u-r$,$M_{r}$ plane
where $M_{r}$ is the Petrosian absolute red magnitude. But
the tight correlation between $M_{*}$ and $M_{r}$  implies
that $M_{r}$ is an accurate indicator of stellar mass.},
($D_{n}(4000)$,$M_{*}$), and 
(H$\delta_{A}$, $M_{*}$) planes.
Specifically, galaxies in the 
($u-r$,$M_{*}$) plane divide into separate blue and red sequences.
Within each sequence,
$u-r$ is correlated with $M_{*}$.
Bimodality in the $u-r$ distribution
arises from extracting the  colors of galaxies within
bins of  constant $M_{*}$.
As $M_{*}$ increases from  
low values, where 
the color distribution is dominated by the blue peak, a red peak grows
while the blue peak declines
in strength until at the highest masses the distribution is dominated
by the red peak. Similar behavior is seen in the ($D_{n}(4000)$,$M_{*}$)
and 
(H$\delta_{A}$, $M_{*}$) planes, except that 
H$\delta_{A}$ is anti-correlated with $M_{*}$.

In Fig.~{\ref{fig:allvswsi}} we plot
DLA analogues to these bivariate galaxy distributions. 
Following the discussion in $\S$ 3.4 we 
substitute  {\w1526} for $M_{*}$ because of evidence that {\w1526} is
an indicator of dark-matter mass, $M_{\rm DM}$ (Prochaska {\etal}
2007), and since
$M_{*}$ should be proportional to  $M_{\rm DM}$. 
Fig.~{\ref{fig:allvswsi}}a shows an analogous and
unambiguous correlation between [M/H]
and log$_{10}${\w1526}. By analogy with the galaxy
result the {\w1526} bin corresponding to low  values of {\w1526}
(log$_{10}${\w1526}({\AA}) $<$ $-$ 0.8)  produces
an [M/H]  distribution  which peaks at [M/H] $\sim$ 
$-$2.3, while the bin corresponding to 
higher values, i.e.,  log$_{10}${\w1526}({\AA})
$>$ $-$ 0.2, gives
rise to an [M/H] distribution, which peaks at [M/H]
$\sim$ $-$0.7. Although bins at intermediate
values of {\w1526} do not result in clear evidence
for two peaks in the [M/H] distribution, this may be a
consequence of small numbers statistics. Furthermore,
small numbers statistics may be responsible for the absence
of a clear discontinuity between the `blue' and `magenta' distributions
in the ([M/H], log$_{10}${\w1526}) plane.  

\begin{figure}[ht]
\figurenum{12}
\includegraphics[width=3.5in]{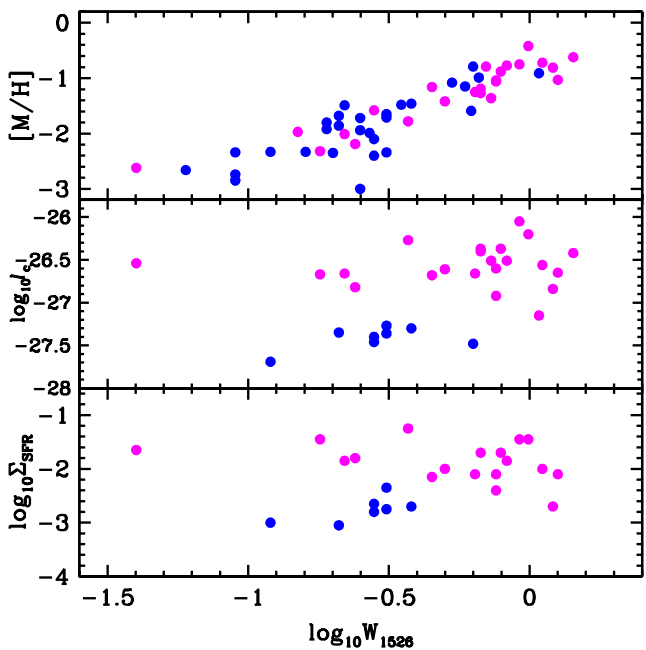}
\caption[]{Bivariate distributions of various parameters
versus {\w1526}. Plots are DLA analogues of bivariate
distributions of various galaxy parameters versus stellar mass.
(a)  Metallicity versus {\w1526}. Blue and magenta
points are `low cool' and `high cool' populations with measured
values of {\w1526}. Note the strong
positive correlation between [M/H] and {\w1526}, and
the dominance of `low cool' points for
low values of {\w1526} and `high cool' points for high values
of {\w1526}.  (b) [{\lc}/{\lcunit}] versus {\w1526}. In this case 
only DLAs with positive detections of {\ciis} are
shown.  The discontinuity between the two populations is obvious but no
evidence for correlations is present. (c) [{\sigmasfr}/{\sfrunit}] 
versus {\w1526}. Systems with high {\sigmasfr}
dominate at large values of {\w1526}. {\it Note, values of {\sigmasfr} for
`high cool' DLAs are lower limits.}
}
\label{fig:allvswsi}
\end{figure}

On the other hand
such a discontinuity is present in
Fig.~{\ref{fig:allvswsi}}b, which plots log$_{10}${\lc}
versus 
log$_{10}${\w1526}. In this case we only plot positive detections,
since the upper and lower limits would obscure possible
trends in the 
({\lc},{\w1526}) plane. While neither
population shows evidence for correlations between
these variables, the bin
with log$_{10}${\w1526}({\AA}) $>$ $-$0.2 
produces an {\lc} disribution dominated by
large cooling rates.  Interestingly, the absence of a clear anti-correlation
between 
log$_{10}${\lc} and log$_{10}${\w1526} runs counter to the
trend in the  (H$\delta_{A}$, $M_{*}$) plane, which
is the closest galaxy analogue, as we have interpreted
{\lc} as an indicator of star formation.
Fig.~{\ref{fig:allvswsi}}c plots log$_{10}${\sigmasfr}
versus log$_{10}$ {\w1526}, which is also analogous
to the bivariate 
distribution in the  (H$\delta_{A}$, $M_{*}$) plane
since {\sigmasfr} is, of course, a star-formation indicator.
In this case, evidence for discontinuities between
the two populations is  strengthened by arguments in $\S$ 4.2,
which 
imply that the
values of {\sigmasfr}  for the
`high cool' DLAs in
Fig.~{\ref{fig:allvswsi}} are conservative lower limits.
We find similar 
results, with somewhat larger scatter,  when {\dv} is  
used as a mass proxy, as shown in 
Fig.~{\ref{fig:allvsdelv90}}.

\begin{figure}
\figurenum{13}
\includegraphics[width=3.5in]{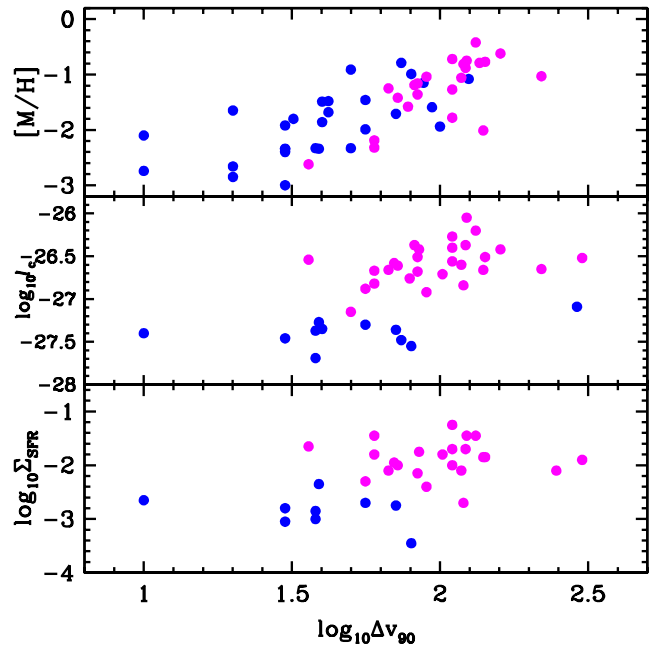}
\caption[]{Same as
Fig.~{\ref{fig:allvswsi}} except that {\dv} is substituted for
{\w1526} as the independent variable.}
\label{fig:allvsdelv90}
\end{figure}

However, we wish to point out an important difference
between the {\ciis} and traditional star-formation 
signatures. 
While the  
H$\delta_{A}$ index is sensitive to star bursts ending
$\sim$ 10$^{8}$ to 10$^{9}$ years before the detection of
H$\delta$ absorption in galaxies, the {\lc} cooling rate
is sensitive only to star formation contemporaneous with
the epoch at which {\ciis} absorption is detected. The reason
for this stems from the short cooling times in DLAs, where

\begin{equation}
  t_{\rm cool}={5 \over 2}{{kT} \over {\ell}_{c}}
\end{equation}

\noindent in the case of CNM gas. Here we have assumed the gas to
be in pressure equilibrium (Wolfire {\etal} 1995)
and we have ignored the contribution
of CMB excitations to the 158 {\micron} emission rate
defined in Eq. 1 (WGP03). Assuming $T$ = 100K, we find cooling
times of $\approx$ 3${\times}10^{6}$ yr and
2${\times}10^{5}$ yr for the median `low cool'
and `high cool' DLAs in Table 3. Therefore, 
our observations are unlikely to detect
158 {\micron} cooling rates from
starbursts with duration
$\Delta t$ short compared to the $\sim$ 2{$\times$}10$^{9}$ yr 
time interval corresponding to the redshift  search window
used to find 
the DLAs in Table 1.
The implication
is that while the star formation history of DLAs may
be  
punctuated with isolated, short-lived starbursts, the {\ciis} technique
is sensitive only to 
a continuous mode of star formation, or multiple
bursts with a high duty cycle. Consequently, we 
may have detected the latter  modes of star formation. 

The short cooling times raise the possibility that the `low cool'
and `high cool' DLAs represent different levels of star formation
activity in the same object. In this scenario an underlying
base level of activity gives rise to the `low cool' DLAs,
while bursts in star-formation rates produce the
`high cool' DLAs. The {\lc} values would
track the star formation rates because of the short cooling times.
However, the metallicity distributions
of the two populations would be indistinguishable since metallicity
is a byproduct of star formation history rather than instantaneous
star formation rate. While the star bursts in the `high cool'
mode could affect the velocity structure of the gas through
stellar winds, the high duty cycle of the bursts implies
that the velocity structure of the gas is also
a function of star-formation history rather than instantaneous
rate. For these reasons it is difficult to understand how
varying star formation rates in a single
class of DLAs could explain  why the
metallicity and velocity structure of the two populations
are so different.

Therefore, in common with modern galaxies the bimodality in DLAs
may arise from the presence of two distinct sequences in mass and
star-formation activity. The obvious question is whether the
DLA phenomenon is the precursor to the bimodality detected in
galaxies? Stated differently, have we detected the early
stages of galaxy bimodality in DLAs?

\subsection{Galaxy Formation Models}

To answer this question we turn to models for the origin
of galaxy bimodality and
then suggest an empirical test.

\subsubsection{Transition Mass Models}

Dekel \& Birnboim (2005) interpret  galaxy bimodality
in terms of a transition  occurring 
at the `shock' dark-matter mass, 
$M_{\rm shock}$$\approx$10$^{11.5}$ {\msolar} (Keres {\etal} 2005;
Dekel \& Birnboim 2005),
corresponding to
$M_{*}$ $\approx$10$^{10.5}$ {\msolar},
which is the stellar mass separating the two sequences
in the (D$_{n}$(4000),$M_{*}$) and (H$\delta_{A}$, $M_{*}$)
planes (Kauffmann {\etal} 2003).
At high redshifts,
halos with {\mdm} $< M_{\rm shock}$ accrete gas in a cold flow
in which the temperature never climbs to the virial temperature
of the halo: such processes result in the formation of disks
observed along the lower-mass blue sequence. By contrast,
gas accreting onto high-redshift halos with 
{\mdm} $> M_{\rm shock}$ is
first heated to the virial temperature ($\sim$10$^{6}$ K)
by stable shocks. The gas later contracts in a quasi-static
spherically symmetric cooling flow to form the 
bulges found in the more massive galaxies populating
the red sequence. In this case outlying cold gas accretes
along
filaments connecting the massive halos to the IGM.

This scenario provides plausible explanations for several
phenomena related to bimodality in DLAs. First, the cold gas
accreting onto halos with {\mdm} $<$ {\mcrit} naturally
evolves into disk-like structures, which act as neutral-gas
reservoirs for {\it in situ} star formation. This is in
accord with our finding that {\it in situ} star formation in 
DLAs is the dominant mode in the `low cool' population, which we identify
with halos having {\mdm} $<$ $M_{\rm shock}$. The low level
of star formation predicted for these objects is probably
related to the low molecular content of the DLA gas  or to
the increase with
redshift of the  Toomre critical surface density (see WC06).
Second, the transition-mass scenario predicts little,
if any, {\it in situ} star formation in the hot gas accreting
onto halos with {\mdm} $>$ {\mcrit}. Rather, star formation in these
objects is predicted to occur in dense, centrally located bulges
that are fueled by filamentary cold flows penetrating the hot gas
(Dekel \& Birnboim 2005; Keres {\etal} 2005). 
This fits in naturally
with our finding that star formation in the `high cool' population,
which we identify with halos having {\mdm} $>$ $M_{\rm shock}$,
mainly occurs in compact star-forming bulges sequestered away from
the DLA gas detected in absorption, which we 
associate with the  dense filamentary
gas predicted in the simulations. This  crucial result 
suggests that the red-blue bimodality in galaxies
and the `high cool'/`low cool' bimodality in DLAs
have a common physical origin. As a result
bimodality in galaxies may originate
in DLAs since it predicts that at large redshifts
the most active star-forming objects are bulges located
in the most massive halos. This is in contrast to 
bimodality in modern galaxies, in which the bulges
and spheroids of massive galaxies exhibit little evidence of
star formation. 
Presumably feedback by AGNs and supernovae that form in the
bulges inhibits star formation  subsequent to the
epochs during which the outlying neutral gas is detected
as high-redshift DLAs (e.g. Governato {\etal} 2004).

If bimodality in DLAs stems from a transition
in star-formation modes at {\mdm}$\approx$ {\mcrit}, then
about half of all DLAs would need to arise in halos with
{\mdm} $>$  {\mcrit}: this conclusion follows
from our finding that about half of our DLA sample is in 
the `high cool' population. Because of the large
value of {\mcrit}, 
the high
median mass $M_{\rm med}$ predicted for DLA  halos would
seem to contradict the standard hierarchical paradigm in
which $M_{\rm med}$ $\approx$ 10$^{10}$ {\msolar}
(Haehnelt {\etal} 1998; Johansson \& Efstathiou 2006). However,
the results of recent numerical simulations reveal a more
complex picture. Nagamine {\etal} (2004) find that 
{\mmed} is a sensitive function of feedback. The model
in best accord with observations is their Q5 run in which 
high-velocity ($\sim$ 400 {\kms}) winds effectively eject
most of the neutral gas from low-mass halos. In this
case {\mmed} $\approx$ 10$^{11.5}$ {\msolar} (Nagamine {\etal}
2007), which agrees with the values predicted
for {\mcrit} and is consistent with the range of
masses deduced from the cross-correlation between
DLAs and LBGs (Cooke {\etal} 2006). While
this model does not explain all the DLA data, 
in particular the predicted area covering factor for
$z$ = [2.5,3.5] is about
a factor of 2 lower than observed, the 
overall
agreement is sufficient to  imply  that
such large values of {\mmed} are plausible.

\begin{figure}[ht]
\figurenum{14}
\includegraphics[height=3.5in,angle=-90]{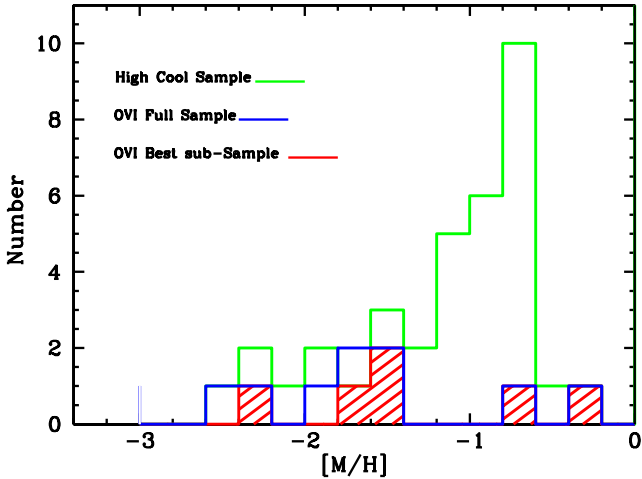}
\caption[]{Comparison between [M/H] histograms for
DLAs in `high cool' sample (green) with blue histogram
depicting 9  DLAs reported to exhibit intervening  O VI absorption
and red striped subsample showing the 6 of these with robust detections
of O VI absorption.}
\label{fig:ovihist}
\end{figure}

\subsubsection{O VI Absorption Tests}

An important feature of the transition-mass model is the presence
of hot gas behind the accretion shock in halos with {\mdm} $>$ {\mcrit}. 
Hot gas is predicted only in this mass range, as halos 
with 
{\mdm} $<$ {\mcrit} accrete gas in cold flows even when heating
due to supernova feedback is included (Keres {\etal} 2005). 
Consequently, hot gas may be a signature of the `high cool'
population. The recent
detection of O VI absorption in a large
fraction of DLAs by Fox {\etal} (2007) 
and their interpretation that O VI is collisionally ionized 
suggests that hot
gas is present in many DLAs. 
Although Fox {\etal} (2007) suggest this gas is
shock heated by supernova remnants, the upper limits on star formation
in DLAs set by WC06 restrict this form of energy
input, provided the WC06 limits apply to gas surrounding
compact LBGs. While such limits
do not yet exist, let us assume that  O VI
absorption in DLAs arises in gas heated by accretion
shocks. 
In that
case
the DLAs with O VI
absorption would belong to the `high-cool' population. 
Searches for {\ciis} absorption  
have been carried out for
three of the six DLAs in which O VI absorption was 
unambiguously detected
in intervening systems. In
DLA0112$-$306 at $z$=2.702
(Srianand {\etal} 2005), {\lc}  exceeds {\lccrit}, while in
DLA2138$-$444 at $z$ = 2.852 (Srianand {\etal} 2005) and 
DLA2206$-$199 at $z$ = 2.076 (see Table 1)  
{\lc} is less than {\lccrit}. Thus, the {\ciis}
data do not provide a statistically meaningful test.

We have considered an alternative test in which
we compared the [M/H] distributions of
DLAs in the `high cool' population with those exhibiting
O VI absorption. The results are shown in
Fig.~{\ref{fig:ovihist}}. When all nine intervening
systems are included in the O VI sample, the KS
tests yields $P_{\rm KS}$([M/H])=0.009 
for the probability that the two
distributions are drawn from the same
parent population. This suggests that 
O VI absorption need not occur in  `high cool' DLAs. 
On the other hand, three of the O VI
identifications are problematic: in each case only
one member of the O IV${\lambda}{\lambda}$ 1031, 1033
doublet was identified, possibly  due to blending of
the missing O VI line with {\lya}
forest absorption lines. But, it is equally plausible that 
all three features were misidentified as O VI. In that
case only six DLAs would exhibit O VI absorption
(see Fig.~{\ref{fig:ovihist}}), and a revised KS tests yields
$P_{\rm KS}$([M/H])=0.15. As a result, the data are not inconsistent with
the hypothesis that the 
`high cool' DLAs are in massive halos.

\section{CONCLUSIONS}

In this paper we searched for evidence of bimodality
in DLAs. Our principal diagnostic tool was the 
[C II] 158 {\micron} cooling rate per H atom, {\lc},
which we obtained from measurements of the {\ciis} $\lambda$ 1335.7
and damped {\lya} absorption lines. In
addition to the {\ciis} transition we used  accurate velocity profiles
of resonance transitions 
to measure velocity widths, equivalent widths, abundances, and
dust-to-gas ratios for 76 DLAs. Our results are
summarized as follows.

(1) Our studies of the {\ciis} $\lambda$ 1335.7 absorption 
line resulted in 37 positive detections, 7 lower limits, and 
32 upper limits of {\lc}. The positive detections show
strong evidence for a bimodal distribution with two peaks
at {\lc}=10$^{-27.4}$ and 10$^{-26.6}$ {\lcunit} separated by
a trough at {\lccrit}$\approx$10$^{-27.0}$ {\lcunit}. In 
$\S$ 2 we argued that the distribution of the
true values of {\lc} corresponding
to the lower and upper limits is consistent with the
distribution of positive detections.  
Stated differently, there is compelling
evidence that the  {\lc} distribution in 
Fig.~{\ref{fig:lchist_pd}}  is a faithful representation of
the parent population from which all the true values of {\lc}
are selected. 

(2) In $\S$ 3 we tested the bimodality hypothesis by the following
method. We first divided the full DLA sample into a `low cool'
subsample with {\lc} $\le$ {\lccrit} and a `high cool' subsample
with {\lc} $>$ {\lccrit}. We then compared distributions of various
physical parameters in the two subsamples. The probability that
the parameters {\dv}, [M/H], $\kappa$, and {\w1526} in
the two subsamples are drawn from
the same parent populations is small, ranging between 10$^{-5}$ to
$\approx$5{$\times$}10$^{-3}$. On the other hand the distributions
of {\nh} are consistent with being drawn from the same parent population.
Therefore, the two subsamples are likely to be separate
populations with  
distinct distributions of velocity width, 
metal abundances, and gas-to-dust ratio, but
similar H I column-density distributions.

(3) In $\S$ 4 we considered different physical processes responsible
for heating the `low cool' and `high cool' DLAs. We studied 
the possibility that the `low cool' DLAs were comprised of warm neutral-medium
gas heated by background radiation alone, but considered 
it to be unlikely because background heating
does not provide sufficient energy input to account for the
cooling rates in the majority of these DLAs.

(4) We then investigated whether
local heating could account for the cooling rates in both DLA populations.
Applying the two-phase model of WPG03, in which FUV radiation
emitted by massive stars heats the gas by the grain photoelectric
mechanism 
and assuming that {\ciis} absorption arises in
CNM gas, we found the resulting SFRs per unit area, {\sigmasfr},
to be bimodal. We further found that the infrequent occurrence
of extended low surface-brightness galaxies in the UDF (WC06)
rules out {\it in situ} star formation as the heat source
for the `high cool' DLAs. 
Rather,
these DLAs are likely heated by FUV radiation emitted by compact
bulge sources, identified as Lyman Break Galaxies, embedded in the neutral gas.
On the other hand {\it in situ} star formation in the `low cool' DLAs
is compatible with the UDF results, provided the DLA diameters
are less than $\sim$ 10 kpc.
The short cooling times of the gas further imply 
that the star formation we detect occurs continuously
rather than in a few isolated bursts.
We conclude that  star formation occurs in all DLAs and
proceeds
in one of two modes, {\it in situ} or bulge dominated. 

(5) In $\S$ 5 we compared the median values of properties
in the two populations. The results suggest that
the crucial parameter distinguishing  the
two populations is mass. That is,
`high cool' DLAs are embedded in halos that are significantly
more massive than the halos encompassing the `low cool' DLAs.

(6) The idea of mass sequences brings up analogies with 
bivariate distributions in modern galaxies.
Recent surveys
reveal  separate
parallel
`blue' and `red' galaxy sequences in which galaxy color and age
are correlated with stellar mass, while star-formation activity is
anti-correlated with stellar mass.
When the galaxy sample is divided into
bins of stellar mass, the color distribution in
each mass bin is bimodal, with an increase in the ratio of
`red' to `blue' galaxies with stellar mass.   
In $\S$ 6
we constructed bivariate
distributions for analogous properties of DLAs. 
Using
{\w1526} (or {\dv}) as  a proxy for mass and 
[M/H], as a proxy for $u-r$ color, we found
strong evidence for a correlation between [M/H] and {\w1526}. 
We also found tentative evidence for two parallel 
sequences in  the ({\lc},{\w1526}) and ({\sigmasfr},{\w1526}) planes,
where {\lc} and {\sigmasfr} are possible proxies for H$\delta_{\rm A}$,
an indicator of recent star formation in galaxies.
We found tentative evidence 
for (1) bimodal distributions
of {\lc} and {\sigmasfr} in bins  
of fixed {\w1526}, and (2) a larger fraction
of `high cool' DLAs with high values of
{\lc} and {\sigmasfr} in bins with the largest values of {\w1526}. 
Therefore,
star-formation activity in
DLAs may increase with mass,
which is consistent with periods of high-$z$ star formation in
early-type galaxies.

(7) In $\S$ 6 we placed these results in the context of
current galaxy formation theory. We found that the transition-mass scenario,
introduced to explain
bimodality in modern galaxies, provides a plausible scenario for the
onset of bimodality in high-$z$ DLAs.
In this picture the `high cool' DLAs comprise outlying neutral gas 
filaments penetrating hot, virialized gas that fills halos
with mass, {\mdm} $>$ {\mcrit}($\equiv$10$^{11.5}$
\msolar). The inflow of these cold streams through the
hot gas results in active star formation in compact
bulges,
which we presume are the embedded LBGs that
heat the surrounding neutral gas. On the other hand accretion
of gas onto halos with
{\mdm} $\le$ {\mcrit} only produces cold inflows that result in the
formation of neutral disks. We associate these objects with the
`low cool' DLAs in which the gas is heated
by {\it in situ} star formation. Since independent simulations demonstrate that
half of the DLA population could have {\mdm} $>$ {\mcrit},
this scenario is in accord with the {\lc} statistics.

To summarize, the most significant result of this paper is that 
the bimodality observed in modern 
galaxies may originate in high-$z$ DLAs. 
According to  Dekel \& Birnboim (2006) at high redshift
the modes of star formation bifurcate around halos with
{\mdm} $\approx$ $M_{\rm shock}$: halos with 
{\mdm} $>$ $M_{\rm shock}$ accrete gas in hot spherical
inflows that are
penetrated by cold streams generating starbursts
in the inner disks or
bulges, 
while the inflow of  cold gas onto
halos with
{\mdm} $<$ $M_{\rm shock}$ leads to disk growth and
{\it in situ} star formation throughout the disks.
This results in a bifurcation in cooling rates, {\lc},
for the following reasons. First, the bifurcation of star
formation modes around $M_{\rm shock}$ creates a bifurcation
in  
$<${\jnulocal}$>$, the locally generated mean intensity
averaged over all impact parameters, which is given
by the same expression for bulge dominated and {\it in situ}
star formation modes (WGP03). In both cases
$<${\jnulocal}$>$  $\propto$ 
{\Mdot}/${\pi}r_{\rm DLA}^{2}$ (Eq. 3), 
where {\Mdot} is the total SFR and $r_{\rm DLA}$ is the radius
of the neutral-gas disk.
While both {\Mdot} and {$r_{\rm DLA}$} presumably depend
on {\mdm}
(Springel \& Hernquist 2003; Mo {\etal} 1998),
they also depend on environmental factors, such as
gas density, which clearly differ for the two populations.
As a result, 
$<${\jnulocal}$>$   
will not be  a continuous function
of {\mdm} near $M_{\rm shock}$, but rather 
will bifurcate. Because {\lc} $\propto$ 
${\kappa}{\epsilon}${\jnulocal} for photoelectrically
heated gas in thermal balance (eq. 2)  and 
the product ${\kappa}{\epsilon}$ is about the same
in the `high cool' and `low cool' DLAs, the {\lc}
distribution should exhibit an analogous bifurcation
or bimodality.
The transition  to  bimodality in  modern galaxies
occurs at lower redshifts, say $z$ $<$ 2, where feedback processes
quench star formation at {\mdm} $>$ $M_{\rm shock}$ (Dekel \& Birnboim
2006) and the
`high cool' DLAs shut down their cooling rates and evolve onto
the red sequence.

Of course, there are caveats to these conclusions.
First, the analogies with mass correlations in modern
galaxies are tentative due to the limited number of DLAs
with positive detections of {\ciis} absorption. We need
to increase the size of the sample to establish the reality
of this result.

The second caveat is related to our assumption that {\dv}
and {\w1526} are measures of virial velocities, hence
of dark-matter mass.
Razoumov {\etal} (2007) describe the difficulties
in constructing self-consistent numerical $\Lambda$CDM models in which
the DLA velocity fields are caused by virial motions alone:
the challenge is to restrict DLA gas to halos sufficiently massive
to generate significant virial velocities and sufficiently
numerous to reproduce the observed DLA covering factors.
An alternative explanation is that gas motions
in DLAs 
are due to outflows; 
i.e. winds (see Nulsen {\etal} 1998; Schaye 2001).
The presence of winds is
plausible in the case of `high cool' DLAs, since  
we have associated these DLAs with  
LBGs, which  exhibit
P Cygni profiles (Pettini {\etal} 2002).
It is reasonable to assume
that DLAs with higher SFRs would produce higher outflow
velocities, which would be consistent with the larger
values of {\dv} and {\w1526} in `high cool' DLAs. 
This possibility is also consistent
with observations of nearby starburst galaxies, which exhibit
correlations between outflow velocities of neutral gas
and  SFRs (Martin 2005).
As a result, the parameters
{\dv} and {\w1526} could be direct signatures of outflows rather
than virial velocities. However, since starburst galaxies
also exhibit a correlation between 
outflow velocity and circular velocity (Martin 2005), the DLA parameters
could be indirect signatures of dark-matter mass even in this scenario.
Therefore, while the physical origin of the velocities
within DLAs  remains controversial, a bimodal 
distribution of dark-matter mass 
is the most plausible explanation for the bimodal properties
of  DLAs.

We finish with the following question: where are the DLAs
in which {\ciis} absorption arises in WNM gas? 
In $\S$ 4.2 we argued that {\ciis} absorption in `high cool'
and `low cool' DLAs arises in
CNM gas exposed, respectively, to high and low values of {\jnulocal}.
But if DLAs are multi-phase media in which the CNM and WNM are in
pressure equilibrium, we would expect
a similar fraction of DLA sightlines to intersect each phase
if, by analogy with the Galaxy ISM, they have comparable
area covering factors (McKee \& Ostriker 1977). Because the grain photoelectric
heating efficiency, $\epsilon$ (Eq. 2), is considerably lower in the WNM
than in the CNM (Weingartner \& Draine 2001), it is also possible for
{\ciis} absorption resulting in {\lc} $\le$ {\lccrit}   
to form in sightlines intersecting {\it only}
WNM gas in `high cool' DLAs. The problem with this scenario 
is that it would also
predict velocity widths, metallicities, and dust-to-gas ratios
significantly larger than observed for the `low cool' DLAs. Thus, while
some  DLAs with {\lc} $\le$ {\lccrit} may  arise in the `high cool'
population,
most of them plausibly form in a separate population
of CNM gas. Therefore, if the area covering factors
of the WNM gas in both populations are significant,
the heating efficiencies
must be sufficiently low for the resulting values of {\lc} to
be less than the 10$^{-27.4}$ {\lcunit} peak of the `low cool'
population. Consequently, the answer to our
question may be that the sightlines encountering only
WNM gas may be responsible for many of the upper limits in
Fig.~{\ref{fig:lcv_NHI}}.

We wish to thank Eric Gawiser, Kim Griest, and Crystal Martin 
for valuable discussions and Jay Strader
for performing the NMIX statistical tests. 
AMW, JXP, and MR were partially supported by 
NSF grant AST 07-09235.   
The W. M. Keck Observatory
is operated as a scientific partnership among the California Institute
of Technology, the University of California and the National Aeronautics and
Space Administration.  The Observatory was made possible by the generous
financial support of the W. M. Keck Foundation.  The authors wish to recognize and
acknowledge the very significant cultural role and reverence that the summit
of Mauna Kea has always had within the indigenous Hawaiian community.  We are
most fortunate to have the opportunity to conduct observations from this
mountain.

\newpage


\newpage



\clearpage

\end{document}